\documentclass[11pt,a4paper]{article}
\pdfoutput=1
\usepackage{jheppub}
\usepackage{gensymb}
\usepackage{placeins}
\usepackage{subfigure}
\usepackage{amssymb,amsmath}
\usepackage{graphicx}
\usepackage{color}
\usepackage{cancel}
\usepackage[colorlinks=true
,urlcolor=blue
,citecolor=blue
,linkcolor=blue
,pagecolor=blue
,linktocpage=true
,pdfproducer=medialab
]{hyperref}
 \usepackage[numbers]{natbib}
\usepackage{notoccite}
\makeatletter \renewcommand{\@dotsep}{10000} \makeatother
\def\be{\begin{equation}}
\def\ee{\end{equation}}
\def\bea{\begin{eqnarray}}
\def\eea{\end{eqnarray}}
\def\bi{\begin{itemize}}
\def\ei{\end{itemize}}

\usepackage[nodisplayskipstretch]{setspace}

\newcommand{\PS}{SU(4)_{c}\times SU(2)_{L}\times SU(2)_{R}}
\newcommand{\mgut}{M_{{\rm GUT}}}

\DeclareUnicodeCharacter{2212}{-}
\DeclareUnicodeCharacter{202F}{\,}

\begin{document}

\begin{titlepage}
\pagestyle{empty}

\vspace*{0.2in}
\begin{center}
{\Large \bf   Muon $\mathbf {g-2}$, Neutralino Dark Matter and Stau NLSP} \\
\vspace{1cm}
{\bf  Mario E. G\'{o}mez$^{a,}$\footnote{Email: mario.gomez@dfa.uhu.es}, Qaisar Shafi$^{b,}$\footnote{E-mail: qshafi@udel.edu}, Amit Tiwari$^{b,}$\footnote{E-mail: amitiit@udel.edu} and
Cem Salih $\ddot{\rm U}$n$^{a,c,}\hspace{0.05cm}$\footnote{E-mail: cemsalihun@uludag.edu.tr}}
\vspace{0.5cm}

{\it
$^a$ Departamento de Ciencias Integradas y Centro de Estudios Avanzados en F\'{i}sica Matem\'aticas y Computación, Campus del Carmen, Universidad de Huelva, Huelva 21071, Spain \\
$^b$Department of Physics and Astronomy,
University of Delaware, Newark, DE 19716, USA \\
$^c$Department of Physics, Bursa Uluda\~{g} University, TR16059 Bursa, Turkey
}

\end{center}

\vspace{0.5cm}
\begin{abstract}

We explore the implications of resolving the muon $g-2$ anomaly in a $SU(4)_c \times SU(2)_L \times SU(2)_R$ model, where the soft supersymmetry breaking scalar and gaugino masses break the left-right (LR) symmetry. A 2 $\sigma$ resolution of the anomaly requires relatively light sleptons, chargino and LSP neutralino. The stau turns out to be the NLSP of mass $m_{\tilde{\tau}}\lesssim 400$ GeV, and the sleptons from the first two families can be as heavy as about 800 GeV. The chargino is also required to be lighter than about 600 GeV to accommodate the muon $g-2$ solutions consistent with the dark matter relic density constraint. The dominant right-handed nature of the light slepton states suppress the sensitivity of possible signals which can be probed in Run3 experiments at the LHC. We also discuss the impact of accomodating the Higgs boson mass and the vacuum stability of the scalar potential for these solutions. The Higgsinos are heavier than about 4 TeV, and the LSP neutralino has the correct relic density if it is Bino-like. We identify stau-neutralino coannihilation as the dominant mechanism for realizing the desired dark matter relic density, with sneutrino-neutralino coannihiliation playing a minor role. These bino-like dark matter solutions can yield a spin-independent scattering cross-section on the order of $10^{-3}$pb which hopefully, can be expected to be tested in the near future.

\end{abstract}
\end{titlepage}


\section{Introduction}
\label{sec:intro}
Supersymmetry (SUSY) is one of the leading candidates {for} exploring physics beyond the Standard Model (SM) since its minimal version, the minimal Supersymmetric SM (MSSM), provides resolution of some of the fundamental challenges faced by the SM. These include {a resolution} of the gauge hierarchy problem, presence of non-baryonic dark matter (DM) candidates with the lightest neutral supersymmetric particle (LSP) being stable by R-parity conservation, {and} stabilization of the electroweak vacuum. In addition, unification of the SM gauge couplings is realized at $\mgut\approx 2 \times 10^{16}$ GeV. The low energy implications of SUSY grand unified theories (GUTs) can be explored in greater detail in experimental observations once the symmetry structure is proposed and the soft supersymmetry breaking (SSB) boundary conditions are imposed at $\mgut$.

SUSY models can also provide a solution to another long standing problem in SM, namely the discrepancy between the theoretical {SM} calculations and the experimental measurements of the muon anomalous magnetic moment (hereafter muon $g-2$). The FermiLab experiment has recently provided a new experimental value for muon $g-2$, which deviates from the SM prediction by $3.3\sigma$ \cite{Abi:2021gix}. This discrepancy between the SM prediction and experiment becomes even larger when combined with the earlier measurements at the Brookhaven National Laboratory \cite{Bennett:2006fi}, leading to the following world average in the muon $g-2$ measurements:

\begin{equation}
\Delta a_{\mu} \equiv a_{\mu}^{{\rm exp}}-a_{\mu}^{{\rm SM}} = (25.1\pm 5.9)\times 10^{-10}~.
\end{equation}

This new world average points to a $4.2\sigma$ deviation from the SM predictions \cite{Davier:2010nc,Hagiwara:2011af,Borsanyi:2020mff}. The supersymmetric contributions to muon $g-2$ arise from a number of super partners which couple to muon at tree-level \cite{Chakraborti:2021dli,Baer:2021aax,Aboubrahim:2021xfi,Wang:2021bcx,Han:2020exx,Altin:2017sxx,Li:2021pnt,Ellis:2021zmg,Athron:2021iuf,Chakraborti:2021bmv,Endo:2021zal,Iwamoto:2021aaf,Baum:2021qzx,Frank:2021nkq,Heinemeyer:2021opc}. In order to solve this discrepancy, the SUSY electroweak sector should involve relatively light neutralinos, charginos and sleptons. On the other hand, the absence of a direct accelerator signal of SUSY raises the experimental bounds on the sparticle mass spectrum. The lower mass bounds on the colored sparticles currently exclude gluinos and squarks of the first two families lighter than about 2 TeV \cite{Vami:2019slp,Aad:2019ftg}. Despite the relatively lower bounds on the electroweak sector, the sensitivity in the current collider experiments is enough to test chargino masses up to about 1.1 TeV and sleptons to about 350 GeV \cite{CMS:2017fdz,Sirunyan:2017zss}. However, these bounds can be loosened if the lighter sleptons are mostly right-handed \cite{Shafi:2021jcg}.

Even though the colored sparticles do not directly contribute to muon $g-2$, the severe bounds on their masses constrain the other particles in SUSY GUTs if universal SSB mass terms are imposed at $\mgut$. For instance, the consistency of Higgs boson mass \cite{ATLAS:2012yve,CMS:2013btf} with the experimental measurements requires stops at the TeV scale \cite{Carena:2011aa}. Therefore, assigning equal masses to superpartners of the matter fields at $\mgut$ implies an increase in the full mass spectrum.

In this work, {we provide a resolution of the muon $g-2$ problem in the framework of supersymmetric $\PS$ ($4-2-2$, for short), without imposing the discrete LR symmetry on the SSB mass terms. In the scalar sector the SSB mass terms break the LR symmetry but are flavor universal.} The paper is organized as follows. We briefly discuss the model and summarize some of its salient features relevant {for} muon $g-2$ and the current experimental results in Section \ref{sec:model}. Section \ref{sec:scan} describes the scanning procedure after a discussion on the experimental constraints employed in our analyses and their impact on the muon $g-2$ resolution. We present our results for the sparticle spectrum compatible with muon $g-2$ in Section \ref{sec:spect}, and {in} Section \ref{sec:DM} {we discuss ways to experimentally test the findings in} the current and near future DM experiments. {We summarize our conclusions} in Section \ref{sec:conc}.

\section{Model Description and Muon $g-2$ confronted with Higgs Boson}
\label{sec:model}

In this section we briefly summarize the supersymmetric $4-2-2$ model \cite{Pati:1974yy,Lazarides:1980tg,Kibble:1982ae} and some of its salient features. Among the many different breaking schemes of $SO(10)$ \cite{Babu:1992ia,Anderson:1993fe,Drees:1986vd,Kawamura:1993uf,Kolda:1995iw,Miller:2012vn,Babu:2005ui}, the $4-2-2$ symmetry can be preserved if $SO(10)$ is broken through the vacuum expectation values (VEVs) of the Higgs fields residing either in $54_{H}$ or $210_{H}$ Higgs multiplets. Even though both breaking mechanisms result in $4-2-2$, they can be distinguished by the presence or absence of the LR symmetry. The breaking through the VEV in $54_{H}$ preserves the so called $C-$parity \cite{Kibble:1982dd,Lazarides:1985my}, which transforms the left-handed and right-handed fields into each other. However, if one implements the $SO(10)$ breaking through the $210_{H}$ VEV, the $C-$parity is broken \cite{Babu:2016bmy}. Note that the SO(10) model involves additional Higgs fields such as $126_{H}$ whose VEV breaks $4-2-2$ to the MSSM gauge symmetry. With this $\overline{126}_{H}$ VEV, the $D-$term contributions to the SSB mass terms are canceled. Furthermore, an unbroken discrete  $\mathcal{Z}_2$ gauge symmetry ensures the stability of the lightest LSP. \cite{Kibble:1982dd}. 

The breaking of $4-2-2$ to the MSSM gauge group leaves intact the hypercharge generator $Y$, where

\begin{equation}
Y = \sqrt{\dfrac{3}{5}} I_{3R} + \sqrt{\dfrac{2}{5}}(B-L)~.
\end{equation}

Here $I_{3R}$ and $B-L$ represent the diagonal generators of $SU(2)_{R}$ and $SU(4)_{c}$, respectively. Taking into account supersymmetry breaking yields the following relation among  the three SSB gaugino masses at the GUT scale:

\begin{equation}
M_{1}=\dfrac{3}{5}M_{2R} + \dfrac{2}{5}M_{4}~,
\label{eq:PSgauginos}
\end{equation}
where $M_{2R}$ and $M_{4}$ denote the gaugino mass terms for $SU(2)_{R}$ and $SU(4)_{c}$, respectively, and $M_{3}=M_{4}$ at $\mgut$. If the LR breaking  in the gaugino sector is parametrised as $M_{2R}=y_{LR}M_{2}$, where $M_{2}$ is the SSB mass  of $SU(2)_{L}$ gaugino, then the SSB gaugino masses become independent. This non-universality in gaugino masses helps to remove the tension between the supersymmetric muon $g-2$ contributions and the fairly severe gluino mass bound.

The broken LR symmetry, in general, implies non-universal masses for the soft left- and right-handed matter scalar masses which can be quantified as $m_{R}\equiv x_{LR}m_{L}$, where $m_{R}$ ($m_{L}$) denotes the SSB mass of the right-handed (left-handed)  fields. In addition to the gaugino and scalar soft masses, the MSSM Higgs fields are realized as superpositions of the appropriate Higgs fields residing in different representations mentioned in the $SO(10)$ and $4-2-2$ breakings. Hence, the boundary conditions for the SSB mass terms also involve non-universal mass terms for the MSSM Higgs fields.

\begin{figure}[htb!]
\centering
\includegraphics[scale=1.8]{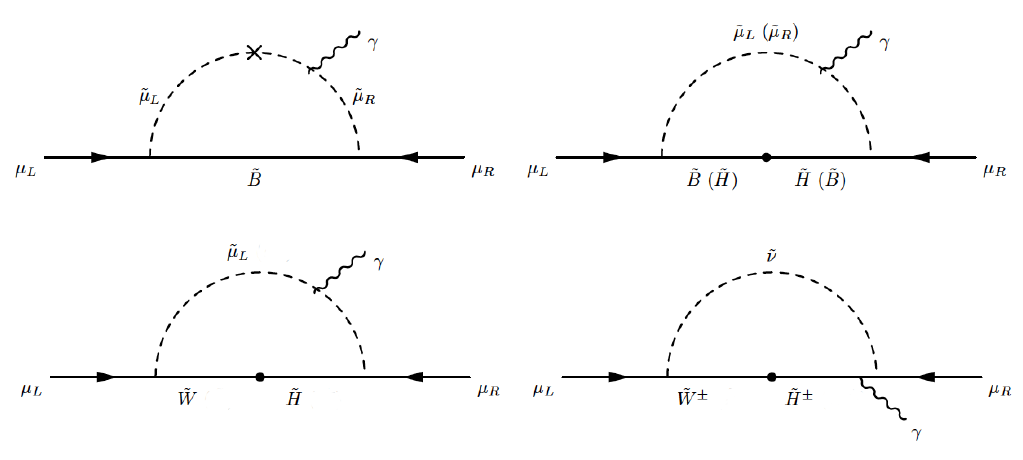}
\caption{The leading contributions to the muon $g-2$ through neutralino and chargino loops. The cross in the top-left diagram denotes the chirality flip between the left- and right-handed smuons, while
the dots in the other diagrams represent the mixing between different Neutralino species. In the top-right diagram, there is another loop which is formed by the particles given in the parentheses.}
\label{fig:SNloops}
\end{figure}

In our work, we assume that the $4-2-2$ symmetry is broken to the MSSM gauge group near the GUT scale, and thus we require the solutions to approximately maintain unification of the SM gauge couplings at $\mgut$. The $4-2-2$ particle spectrum contains the right-handed neutrinos, and the tiny neutrino masses established by the current experiments \cite{Super-Kamiokande:2010orq} require either very heavy right-handed neutrinos or negligible couplings between the MSSM particles and the right-handed neutrinos ($Y_{\nu} \lesssim 10^{-7}$) \cite{Coriano:2014wxa,Khalil:2010zza,Abbas:2007ag}, and so the right-handed neutrinos effectively decouple from the MSSM spectrum. Thus, the model below $\mgut$ turns out to be MSSM. The supersymmetric muon $g-2$ contributions arise from the electroweak sector of MSSM involving the Bino, Wino, Higgsinos and sleptons \cite{Moroi:1995yh,Martin:2001st,Giudice:2012pf}, and the leading one-loop diagrams are shown in Figure \ref{fig:SNloops}. The top-left diagram in Figure \ref{fig:SNloops}, involves the Bino and smuons and it represents the dominant contribution to muon $g-2$, which is enhanced by the chirality flip (shown with a cross) between the smuons. The contribution  is proportional to $\mu\tan\beta$, where $\mu$ represents the bilinear mixing term of the MSSM Higgs fields, and $\tan\beta$ is the ratio of their VEVs, $\tan\beta \equiv v_{u}/v_{d}$. Thus, the solution of the muon $g-2$ discrepancy favors large $\mu\tan\beta$ values, and the dominant contributions to muon $g-2$ can be calculated approximately as \cite{Fargnoli:2013zia}

\begin{equation}
\Delta a_{\mu}^{\tilde{B}\tilde{\mu}_{L}\tilde{\mu}_{R}} \simeq \dfrac{g_{1}^{2}}{16\pi^{2}}\dfrac{m_{\mu}^{2}M_{\tilde{B}}(\mu\tan\beta - A_{\mu})}{m_{\tilde{\mu}_{L}}^{2}m_{\tilde{\mu}_{R}}^{2}}F_{N}\left( \dfrac{m_{\tilde{\mu}_{L}}^{2}}{M_{\tilde{B}}^{2}},\dfrac{m_{\tilde{\mu}_{R}}^{2}}{M_{\tilde{B}}^{2}}\right)~,
\label{eq:binocont}
\end{equation}
where

\begin{equation}
F_{N}(x,y) = xy\left[\dfrac{-3+x +y + xy}{(x-1)^{2}(y-1)^{2}} + \dfrac{2x\log(x)}{(x-y)(x-1)^{3}} -\dfrac{2y\log(y)}{(x-y)(y-1)^{3}}\right]~.
\label{eq:loopfunction}
\end{equation}

The remaining diagrams in Figure \ref{fig:SNloops} also contribute to muon $g-2$ through the mixing of Higgsinos with Bino and Wino, with similar contributions from the sneutrinos. However, these three diagrams are suppressed by the small Yukawa couplings of the lighter generations. In sum, the relevant parameters for this process are listed in Table \ref{tab:fund}.

\begin{table}[h!]
\centering
\begin{tabular}{c|c}
Low Scale & GUT Scale \\ \hline
$m_{\tilde{\mu}_{L}}, m_{\tilde{\nu}}$ & $m_{L}$ \\ 
$m_{\tilde{\mu}_{R}}$ & $m_{R}$ \\
$M_{\tilde{B}}$ & $M_{1}$ \\
$M_{\tilde{W}}$ & $M_{2}$ \\
$\mu$ & $m_{H_{u}}, m_{H_{d}}$ \\
$A_{\mu}$ & $A_{0}$ \\
$\tan\beta$ & $\tan\beta$ \\ \hline
\end{tabular}
\caption{The fundamental parameters determining the supersymmetric contributions to muon $g-2$ at the low scale (left) and GUT scale (right).}
\label{tab:fund}
\end{table}

We emphasize that the non-universality imposed in the SSB scalar masses is family independent and it distinguishes only the left-handed supersymmetric particles from the right-handed ones. In this context, there exists a tension between the desired muon $g-2$ contributions and the measured SM-like Higgs boson mass. The SUSY spectrum needs relatively heavy third family  masses in order to accommodate a Higgs boson mass of about 125 GeV, whereas sizeable contributions to muon $g-2$ favor relatively light sparticles, at least in the electroweak sector. The loop contributions to the Higgs boson mass in the MSSM framework can be written as \cite{Carena:2012mw}

\begin{equation*}
\Delta m_{h}^{2}\simeq \dfrac{m_{t}^{4}}{16\pi^{2}v^{2}\sin^{2}\beta}\dfrac{\mu A_{t}}{M^{2}_{{\rm SUSY}}}\left[\dfrac{A_{t}^{2}}{M^{2}_{{\rm SUSY}}}-6 \right]+
\end{equation*}
\begin{equation}\hspace{1.4cm}
\dfrac{y_{b}^{4}v^{2}}{16\pi^{2}}\sin^{2}\beta\dfrac{\mu^{3}A_{b}}{M^{4}_{{\rm SUSY}}}+\dfrac{y_{\tau}^{4}v^{2}}{48\pi^{2}}\sin^{2}\beta \dfrac{\mu^{3}A_{\tau}}{m_{\tilde{\tau}}^{4}}~~,
\label{eq:higgscor}
\end{equation}
where the first line represents the contributions from the stops, and contributions depicted in the second line come from the sbottom and stau, respectively.

The stops contribution is dominant in Eq.~(\ref{eq:higgscor}), whereas the contributions from the sbottom and stau can be important for moderate and large $\tan\beta$ values. However, the vacuum stability constraints on $\mu\tan\beta$  \cite{Carena:2012mw,Hisano:2010re,Kitahara:2013lfa} only allows minor contributions from these terms. Nevertheless, the solutions with moderate or large $\tan\beta$ also yield some suppression in the contributions from the stop. Thus, such acceptable solutions need heavy stops and/or large trilinear coupling ($A_{t}$) to satisfy the Higgs boson mass constraint. On the other hand, as one can see from Eq.(\ref{eq:binocont}), the solutions with moderate or large $\tan\beta$ can give rise to a  parameter space compatible with the muon $g-2$ solution since the supersymmetric contributions are enhanced by $\tan\beta$. Therefore, if family-independent SSB mass terms are imposed at the GUT scale, the sparticles from the first two families become sensitive to the impact from the Higgs boson mass on the third family sparticles. One may have to compromise to accommodate the muon $g-2$ resolution with a consistent Higgs boson mass.

Another tension between the Higgs boson mass and the muon $g-2$ resolution arises if one considers the vacuum stability of the MSSM scalar potential. Resolving  muon $g-2$ yields relatively light staus in the spectrum compatible with the Higgs boson mass constraint, but the negative coupling between the Higgs fields and the staus tends to destabilize the vacuum, and the metastability condition on the MSSM scalar potential can have a strong impact for stau masses lighter than about 1 TeV. An overall upper bound from the metastability condition can be expressed as $\mu\tan\beta \lesssim 100$ TeV \cite{Carena:2012mw,Hisano:2010re,Kitahara:2013lfa}, if both the left-handed and right-handed staus are lighter than about 600 GeV. However, this can increase to 700 TeV or so if one of the staus is heavier than about 1 TeV \cite{Shafi:2021jcg}. Hence, one cannot realize arbitrarily large values of $\mu$ and $\tan\beta$ to obtain the desired supersymmetric muon $g-2$ contributions. 

Indeed, the parameter space compatible with the experimental muon $g-2$ values in SUSY GUTs can lead to $\mu$ values up to about 5 TeV or so, even though it is possible to satisfy the vacuum stability constraint with $\mu$ as large as 10 TeV. A large value of $\mu$ would also imply large values for the other SSB masses in the scalar potential. For instance, the minimization of the scalar potential leads to the relation:

\begin{equation}
\dfrac{M_{Z}^{2}}{2} = \dfrac{m_{H_{d}}^{2}-m_{H_{u}}^{2}\tan\beta}{\tan^{2}\beta -1} - \mu^{2}~,
\label{eq:MZ}
\end{equation}
where $m_{H_{d}}$ and $m_{H_{u}}$ are the SSB mass terms for the MSSM Higgs fields, and we assume the loop corrections to be included in these mass terms. Considering the suppression from $\tan\beta$ on $m_{H_{d}}^{2}$, Eq.(\ref{eq:MZ}) can be approximated as $M_{Z}^{2} \approx -2m_{H_{u}}^{2} - 2\mu^{2}$. Thus, realizing large $\mu$  at the low scale requires one to impose either large $M_{1}$ and $M_{2}$ values, or large $m_{H_{u}}$ values at the GUT scale. Large $M_{1}$ and $M_{2}$ values cause a direct suppression in the muon $g-2$ contributions as can be seen from Eqs.(\ref{eq:binocont} and \ref{eq:loopfunction}). Similarly, if one inputs a large $m_{H_{u}}$ value at the GUT scale, it significantly enhances the scalar masses through Renormalization group equations (RGEs) and yields heavy sleptons in the spectrum, which also suppress the muon $g-2$ contributions. Therefore, even though the muon $g-2$ resolution seems to favor large $\mu$ and $\tan\beta$ values, it cannot be easily accommodated in SUSY GUTs with large values for $\mu$ and $\tan\beta$ due to negative impacts on the Higgs boson mass, vacuum stability and RGE flow.

\section{Scanning Procedure and Experimental Constraints}
\label{sec:scan}

We perform random scans in the fundamental parameter space of $4-2-2$ with broken LR symmetry by using SPheno-4.0.4 \cite{Porod:2003um,Porod:2011nf} and SARAH-4.14.4 \cite{Staub:2008uz,Staub:2015iza}. The fundamental parameter space is spanned by the  parameters listed as follows:

\begin{equation}
\begin{array}{lll}
0 \leq & m_{L} & \leq 5 ~{\rm TeV} \\
0 \leq & M_{2L} & \leq 5 ~{\rm TeV} \\
-3 \leq & M_{3} & \leq 5 ~{\rm TeV} \\
-3 \leq & A_{0}/m_{L} & \leq 3 \\
1.2 \leq & \tan\beta & \leq 60 \\
0\leq & {\rm x}_{{\rm LR}} & \leq 3 \\
-3 \leq & {\rm y}_{{\rm LR}} & \leq 3 \\
0 \leq & {\rm x}_{{\rm d}} & \leq 3 \\
-1 \leq & {\rm x}_{{\rm u}} & \leq 2~. \\
\end{array}
\label{eq:paramSpacePSLR}
\end{equation}
Here $x_{d}$ and $x_{u}$ parametrise the non-universality in the SSB masses at $\mgut$ of the MSSM Higgs fields as $m_{H_{d}}^{2} = x_{d}m_{L}^{2}$ and $m_{H_{u}}^{2} = x_{u}m_{L}^{2}$. In addition, we vary the universal trilinear scalar interaction term $A_{0}$ by {requiring that} the magnitude of its ratio to $m_{L}$ {is not} greater than 3 to avoid the color/charge breaking minima of the scalar potential \cite{Ellwanger:1999bv,Camargo-Molina:2013qva,Camargo-Molina:2013sta}. Note that Eq.(\ref{eq:paramSpacePSLR}) does not represent all the phenomenologically available parameter space, but it is rather optimized to emphasize the muon $g-2$ impact on the fundamental parameter space, and the ranges are determined based on the discussion in Section \ref{sec:model}.

All solutions in our scans are required to be compatible with the unification condition on the SM gauge couplings at $\mgut$. The SPheno package first runs RGEs for the gauge and Yukawa couplings from $M_{Z}$ to $\mgut$, which is calculated as a scale at which the unification condition is realized as $g_{1}=g_{2}\simeq g_{3}$, and the deviation of $g_{3}$ from unification is restricted not to be larger than about $3\%$. After $\mgut$ is determined, the RGEs are run back from $\mgut$ to $M_{Z}$ together with the SSB terms determined by the parameters listed in Eq.(\ref{eq:paramSpacePSLR}). The results after the RGE evolution involve the low scale mass spectrum, mixings, decay channels and their branching ratios. In calculating the DM observables, we require one of the MSSM neutralinos to be the LSP for all the solutions, which is a compelling DM candidate. The SPheno outputs are transferred to micrOMEGAs \cite{Belanger:2018ccd}, which calculates the relic density and the scattering cross-sections of the DM candidates.

After generating the data, we successively apply the mass bounds \cite{Agashe:2014kda} and the constraints from rare $B-$meson decays \cite{Aaij:2012nna,Amhis:2012bh} as well as the current results from the Planck satellite \cite{Akrami:2018vks} on the relic dark matter density. The constraints from rare $B-$meson decays yield a strong impact mostly in the MSSM Higgs spectrum. The current experimental results on these decay modes reveal a strong agreement with the SM predictions and so there is little room for new physics contributions. This agreement has been strengthened even further after the observation of $B_{s}\rightarrow \mu^{+}\mu^{-}$\cite{Aaij:2012nna}. These processes are mediated by the CP-odd Higgs boson in MSSM and their rate is proportional to $\tan^{6}\beta/m_{A}^{4}$ \cite{Babu:1999hn,Choudhury:1998ze}, and consequently consistency with these experiments requires a heavy $CP-$odd Higgs boson especially in regions where $\tan\beta$ varies from moderate to large values. Similarly, the constraint on ${\rm BR}(B_{s}\rightarrow X_{s}\gamma)$ leads to a strong lower bound on the charged Higgs boson. 

The DM considerations, following the Planck Satellite measurements, impose severe {restrictions} on SUSY models with a neutralino a DM candidate. Moreover, if the LSP neutralino is mostly Bino-like, models in which the muon $g-2$ problem is resolved typically predict an inconsistently large relic DM density. However, these values become compatible with the Planck satellite measurements in models with the appropriate annihilation and LSP-NLSP (next to LSP) coannihilation channels.

As discussed before, a resolution of the muon $g-2$ problem favors a light mass spectrum for the electroweak sector particles which may yield multiple coannihilation scenarios. In this context, the NLSP and its interactions with the LSP play an important role in the DM phenomenology. It is also possible to realize a Higgsino-like DM neutralino, and in such cases the relic density constraint excludes the LSP mass up to about 700 GeV. The Higgsino-like LSP neutralino is also constrained by the direct DM detection experiments (see, for instance, \cite{Raza:2018jnh}). A similar discussion also holds for the Wino-like LSP neutralino. We can summarize the experimental constraints employed in our analyses as follows:

\begin{equation}
\setstretch{1.8}
\begin{array}{c}
m_h  = 123-127~{\rm GeV}\\
m_{\tilde{g}} \geq 2.1~{\rm TeV}~(800~{\rm GeV}~{\rm if~it~is~NLSP})\\
0.8\times 10^{-9} \leq{\rm BR}(B_s \rightarrow \mu^+ \mu^-) \leq 6.2 \times10^{-9} \;(2\sigma) \\
2.99 \times 10^{-4} \leq  {\rm BR}(B \rightarrow X_{s} \gamma)  \leq 3.87 \times 10^{-4} \; (2\sigma) \\
0.114 \leq \Omega_{{\rm CDM}}h^{2} \leq 0.126~.
\label{eq:constraints}
\end{array}
\end{equation}
Note that although we have listed only the mass bounds on the Higgs boson and gluino because they have been model independently updated by the Large Hadron Collider (LHC) analyses, we employ the model independent bounds such as those from the Linear electron-positron collider (LEP2), which exclude any new charged particle lighter than about 100 GeV. Despite the precise experimental determination of the Higgs boson we allow an uncertainty of about 2 GeV, which accounts for the theoretical uncertainties. We set the top quark mass to its central value ($m_{t}=173.3$ GeV \cite{CDF:2009pxd}). Even though the supersymmetric spectrum is not very sensitive to the top quark mass, $1\sigma$ or $2\sigma$ variation in its mass together with the uncertainties in the strong gauge coupling yields a $2-3$ GeV shift in the SM-like Higgs boson mass \cite{Gogoladze:2011aa,AdeelAjaib:2013dnf,Degrassi:2002fi}. The supersymmetric particles yield about 1.5 (0.5) GeV uncertainty in the Higgs boson mass in the case of large (small) stop mixing \cite{Bahl:2019hmm,Bagnaschi:2017xid, Athron:2016fuq,Allanach:2004rh,Drechsel:2016htw}. 

Of course, we require  the solutions to be compatible with radiative electroweak symmetry breaking (REWSB), which constrains $\mu$ with respect to the SSB mass terms $m_{H_{u}}$ and $m_{H_{d}}$ as given in Eq.(\ref{eq:MZ}). In addition, $\mu$ can be constrained further by the metastability condition on the MSSM scalar potential, which is discussed in Section \ref{sec:model}. However, these constraints can bound only the magnitude of $\mu$ and, following most conventions, we assume it to be positive for all the solutions generated in our scans.

\section{Sparticle Spectrum and Muon $\mathbf{g-2}$}
\label{sec:spect}

\begin{figure}[ht!]
\centering
\subfigure{\includegraphics[scale=0.4]{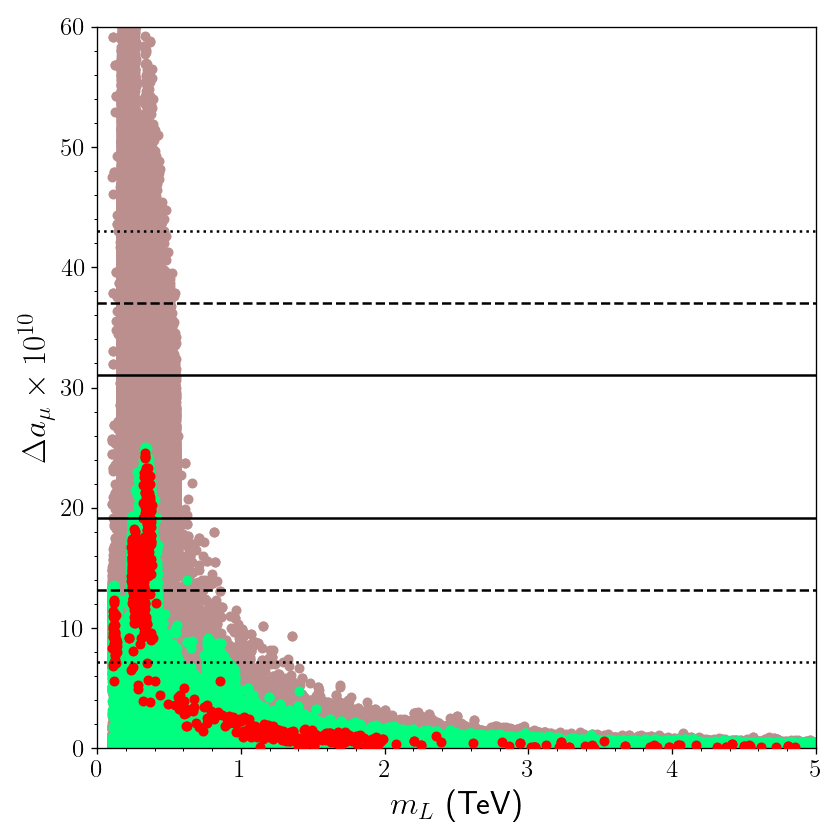}}%
\subfigure{\includegraphics[scale=0.4]{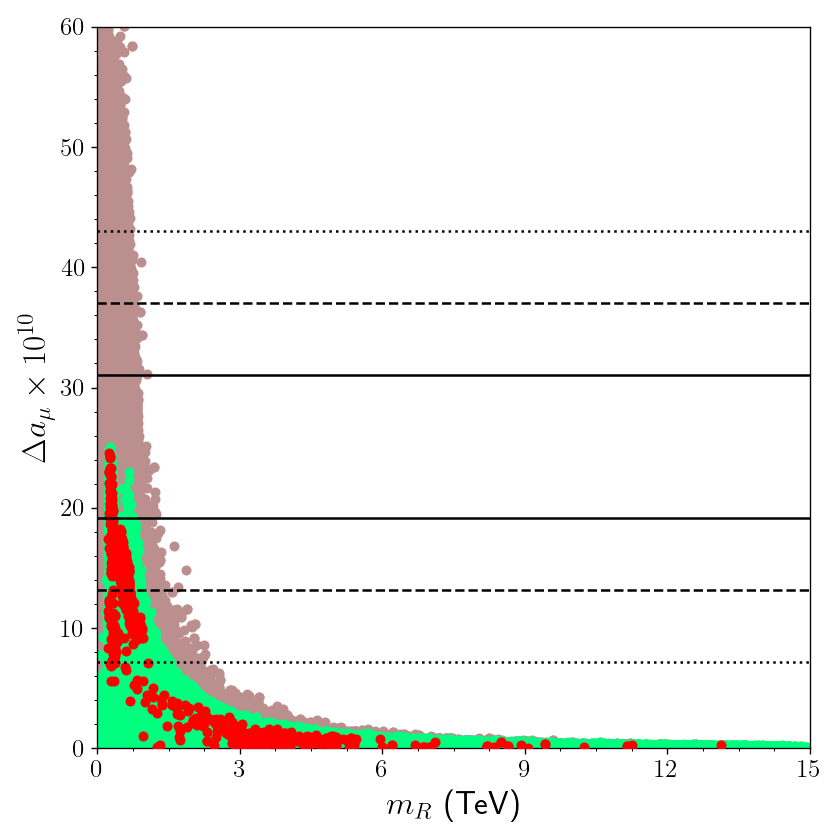}}
\subfigure{\includegraphics[scale=0.4]{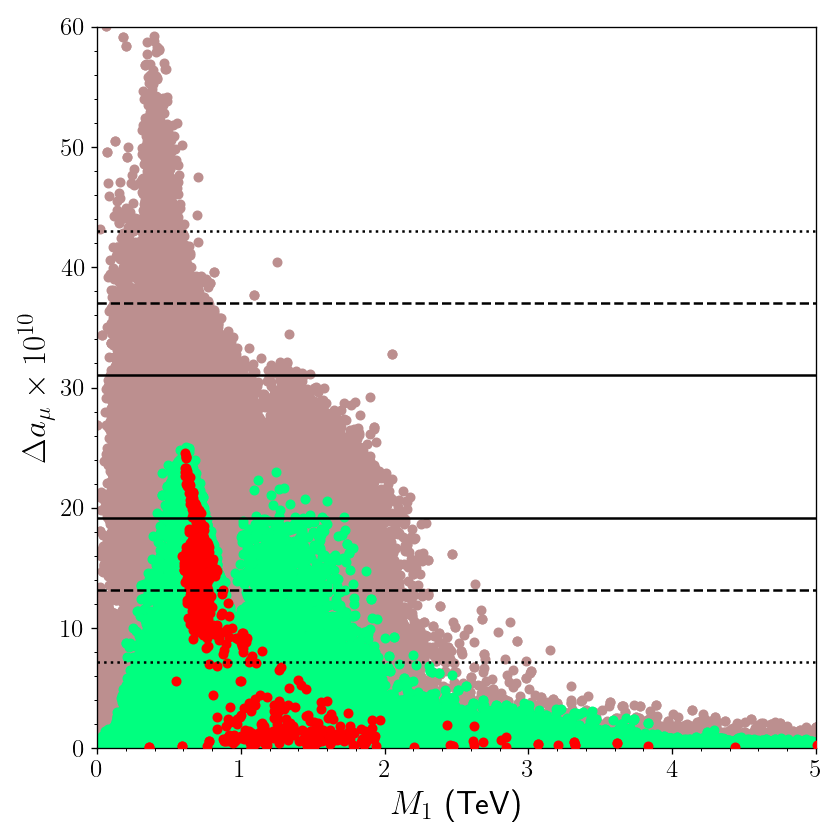}}%
\subfigure{\includegraphics[scale=0.4]{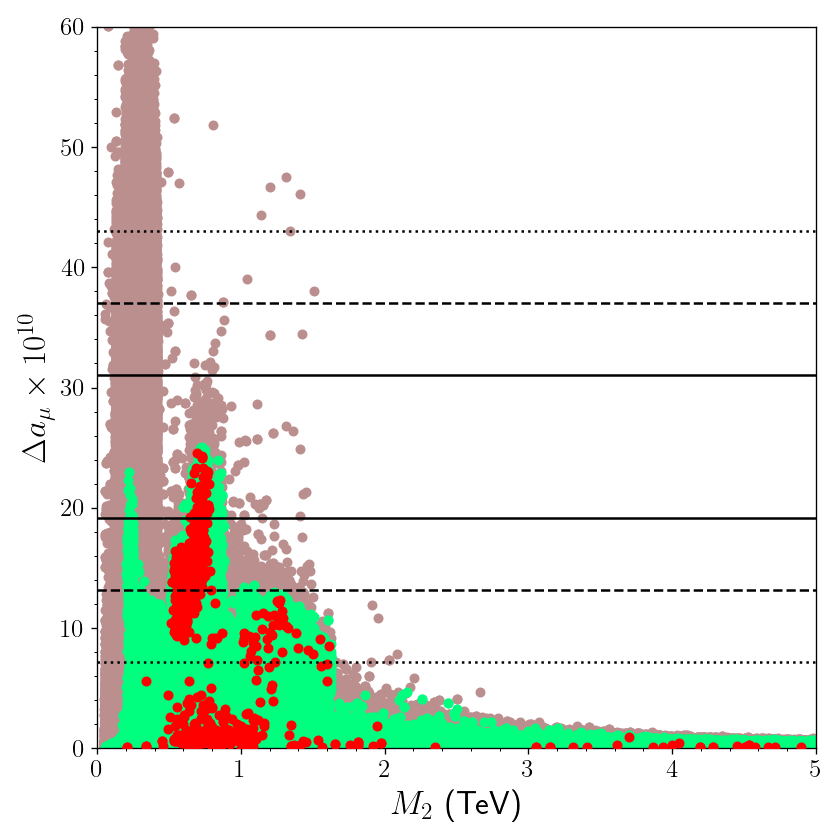}}
\caption{Plots in the $\Delta a_{\mu}-m_{L}$, $\Delta a_{\mu}-m_{R}$, $\Delta a_{\mu}-M_{1}$ and $\Delta a_{\mu}-M_{2}$ planes. All solutions are compatible with the REWSB and LSP neutralino conditions. The green points are allowed by the mass bounds and constraints from rare $B-$meson decays. The red points form a subset of green and they satisfy the Planck bound on the relic density of LSP neutralino within $5\sigma$. The horizontal solid, dashed and dotted lines bound the regions which accommodate the muon $g-2$ resolution within $1\sigma$, $2\sigma$ and $3\sigma$ respectively.}
\label{fig:scalargauginos}
\end{figure}

\begin{figure}[ht!]
\centering
\subfigure{\includegraphics[scale=0.4]{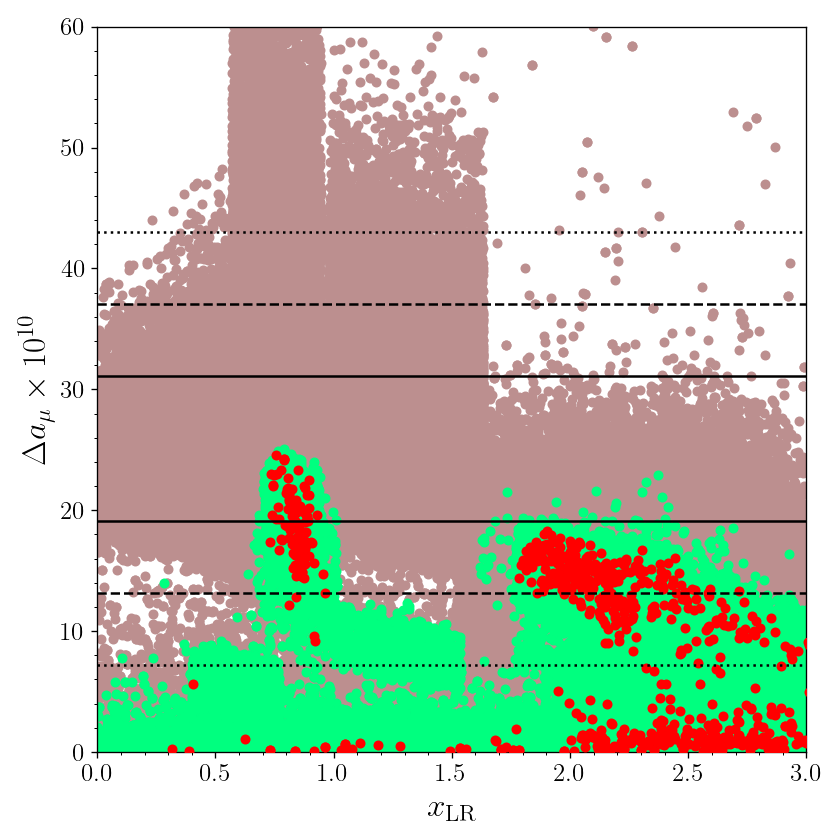}}%
\subfigure{\includegraphics[scale=0.4]{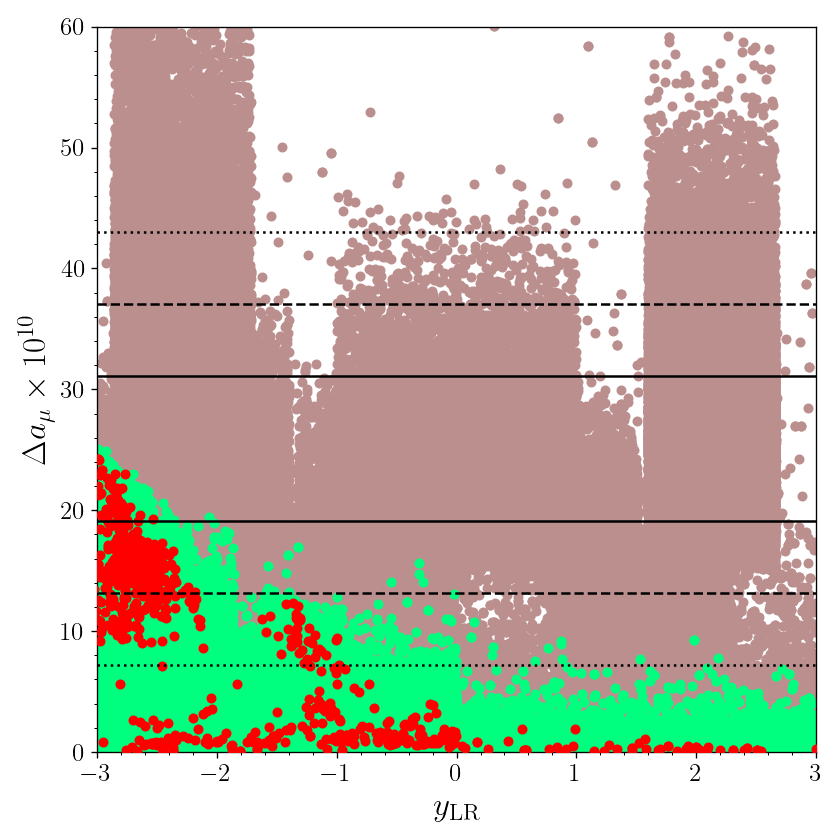}}
\caption{LR breaking and muon $g-2$ implications in the $\Delta a_{\mu}-x_{LR}$ and $\Delta a_{\mu}-y_{LR}$ planes. The meanings of colors and horizontal lines are the same as in Figure \ref{fig:scalargauginos}.}
\label{fig:damuLRbreaking}
\end{figure}

\begin{figure}[ht!]
\centering
\subfigure{\includegraphics[scale=0.4]{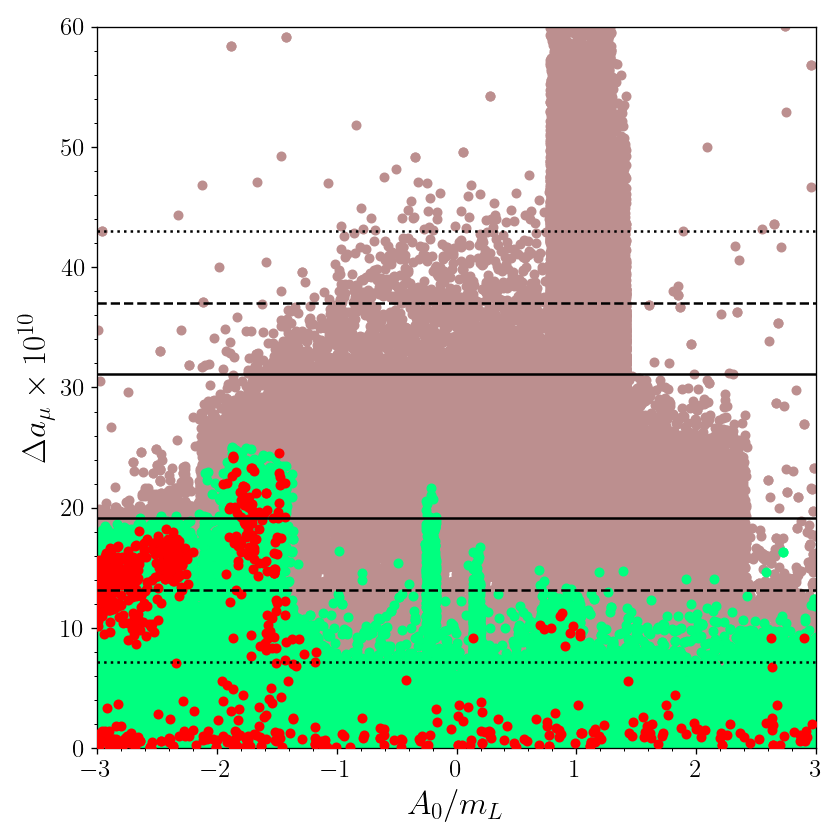}}%
\subfigure{\includegraphics[scale=0.4]{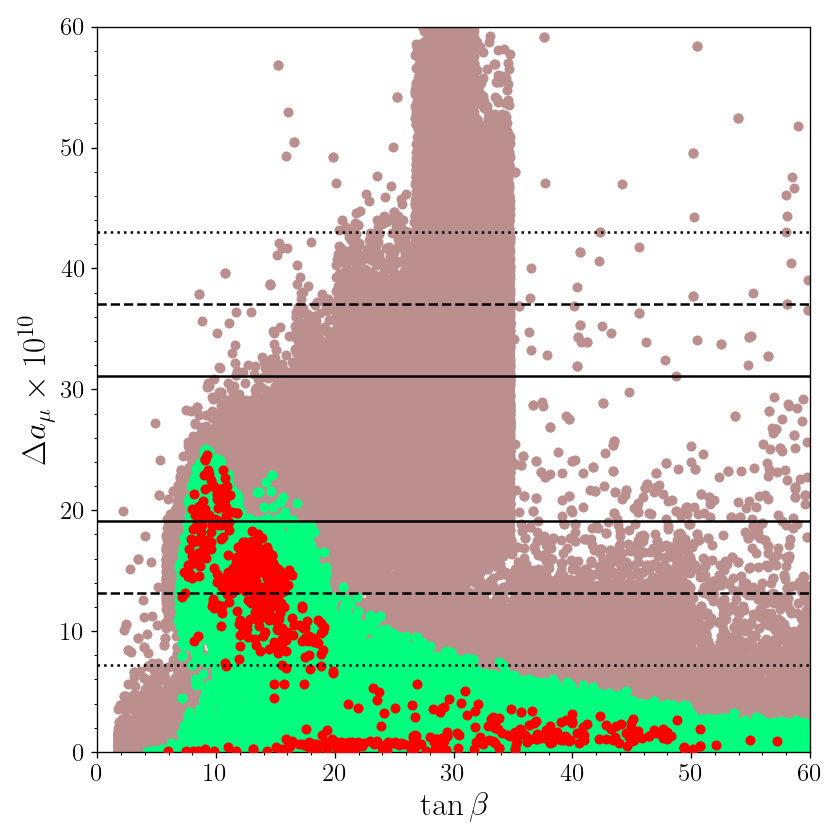}}
\subfigure{\includegraphics[scale=0.4]{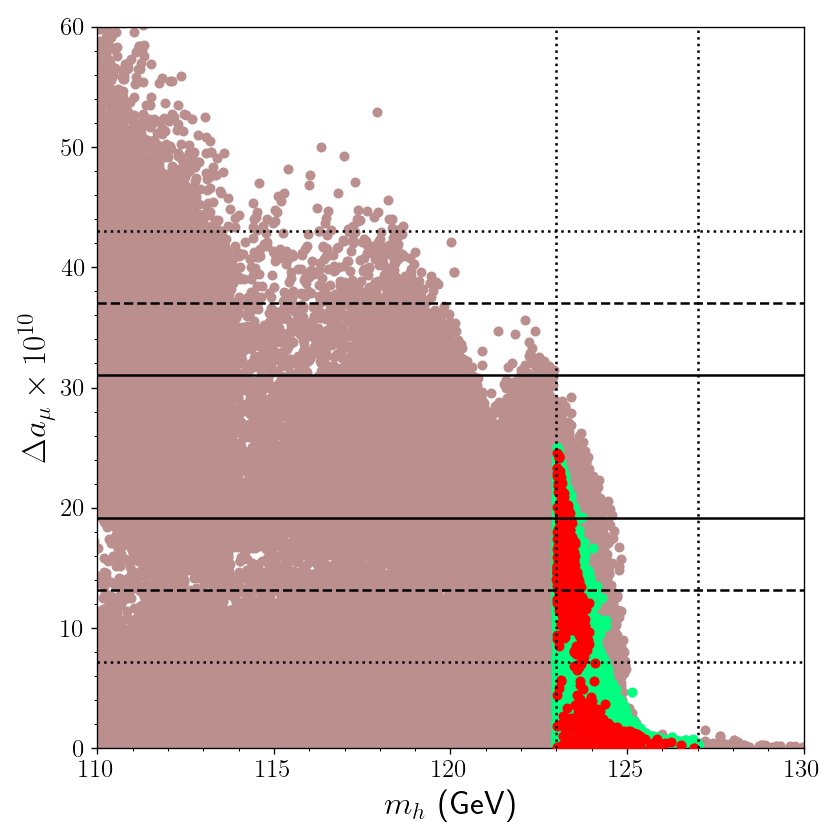}}
\subfigure{\includegraphics[scale=0.4]{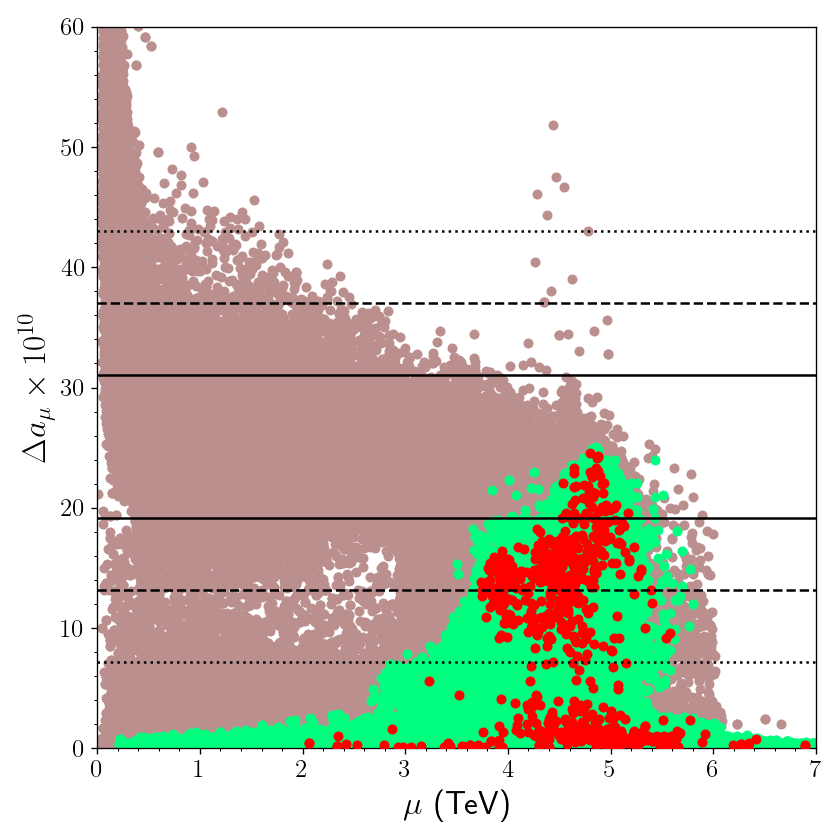}}
\caption{Plots in the $\Delta a_{\mu}-A_{0}/m_{L}$, $\Delta a_{\mu}-\tan\beta$, $\Delta a_{\mu}-m_{h}$ and $\Delta a_{\mu}-\mu$ planes. The meanings of colors and horizontal lines are the same as in Figure \ref{fig:scalargauginos}.}
\label{fig:damuhiggs}
\end{figure}

\begin{figure}[ht!]
\centering
\subfigure{\includegraphics[scale=0.4]{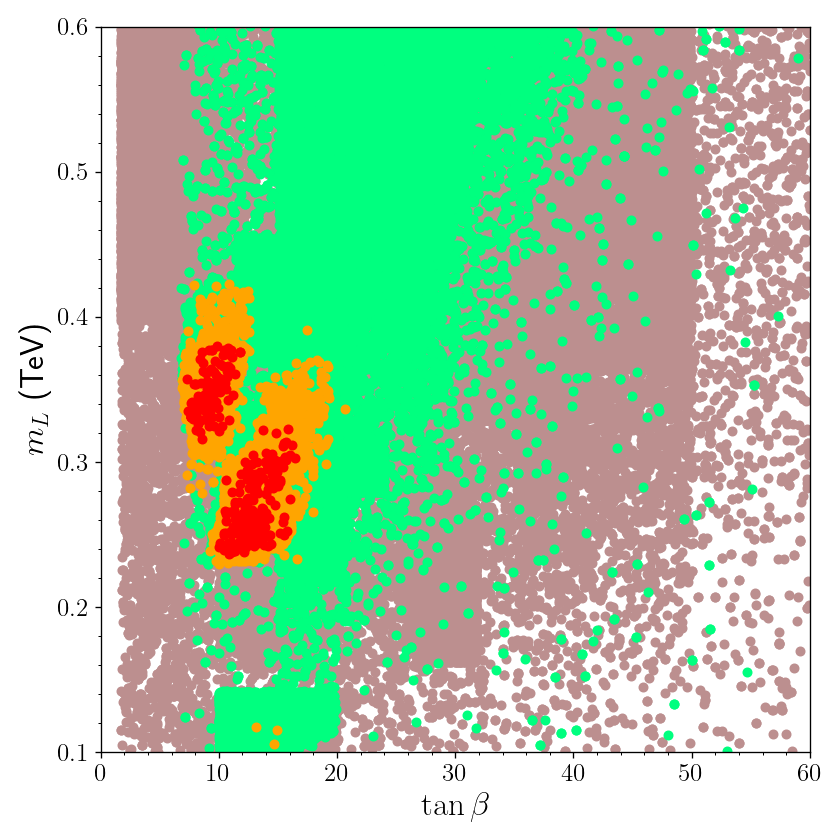}}%
\subfigure{\includegraphics[scale=0.4]{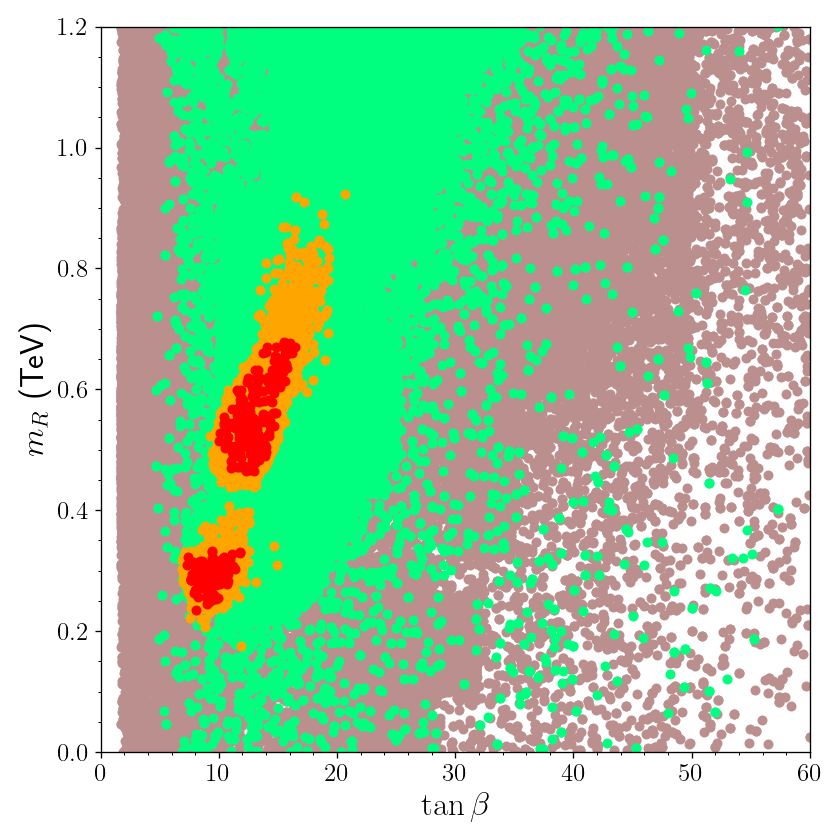}}
\subfigure{\includegraphics[scale=0.4]{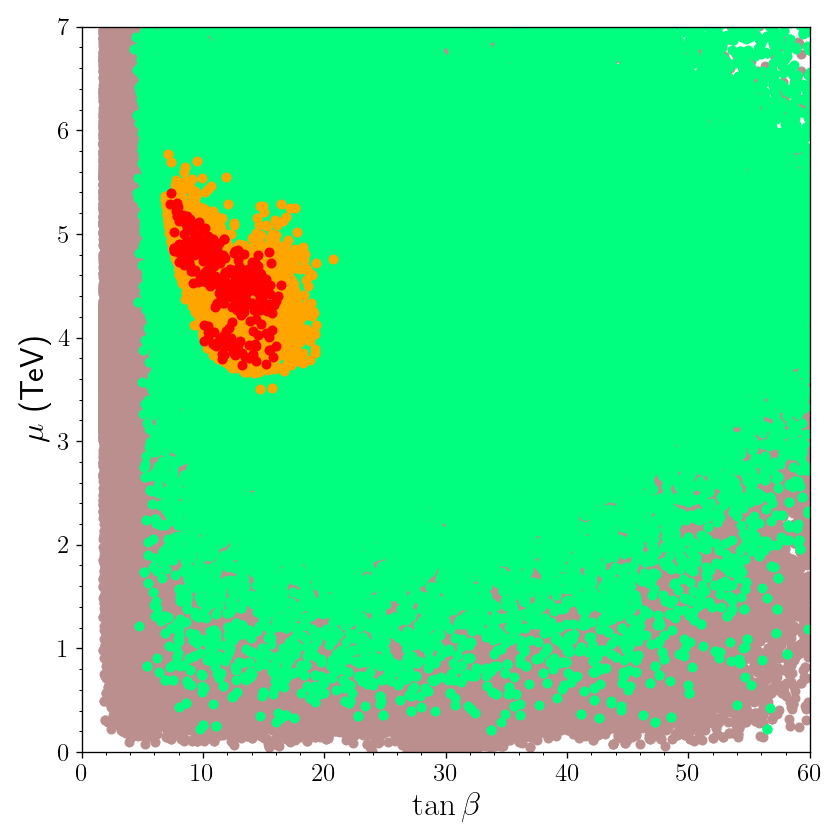}}%
\subfigure{\includegraphics[scale=0.4]{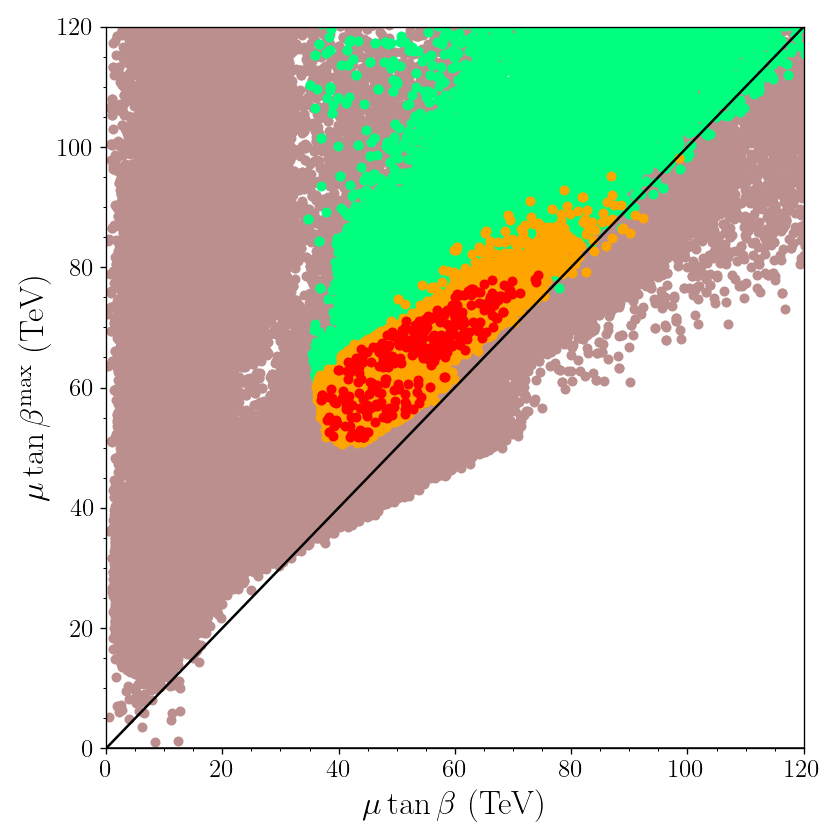}}
\caption{The vacuum stability in $4-2-2$ with plots in the $m_{L}-\tan\beta$, $m_{R}-\tan\beta$, $\mu-\tan\beta$ and $\mu\tan\beta^{{\rm max}}-\mu\tan\beta$ planes. All points are compatible with the REWSB and LSP neutralino conditions. Green points are allowed by the mass bounds and constraints from rare $B-$meson decays. The orange points form a subset of green, and they accommodate the resolution to the muon $g-2$ discrepancy within $2\sigma$. The red points as a subset of orange, additionally yield the correct relic density of the LSP neutralino measured by the Planck satellite within $5\sigma$ uncertainty. $\mu\tan\beta^{{\rm max}}$ represents the possible maximum values of $\mu\tan\beta$ allowed by the metastability condition on the scalar potential, and the the diagonal line in the $\mu\tan\beta^{{\rm max}}-\mu\tan\beta$ plane indicates the solutions for which $\mu\tan\beta = \mu\tan\beta^{{\rm max}}$.}
\label{fig:metastability}
\end{figure}

\begin{figure}[ht!]
\centering
\subfigure{\includegraphics[scale=0.4]{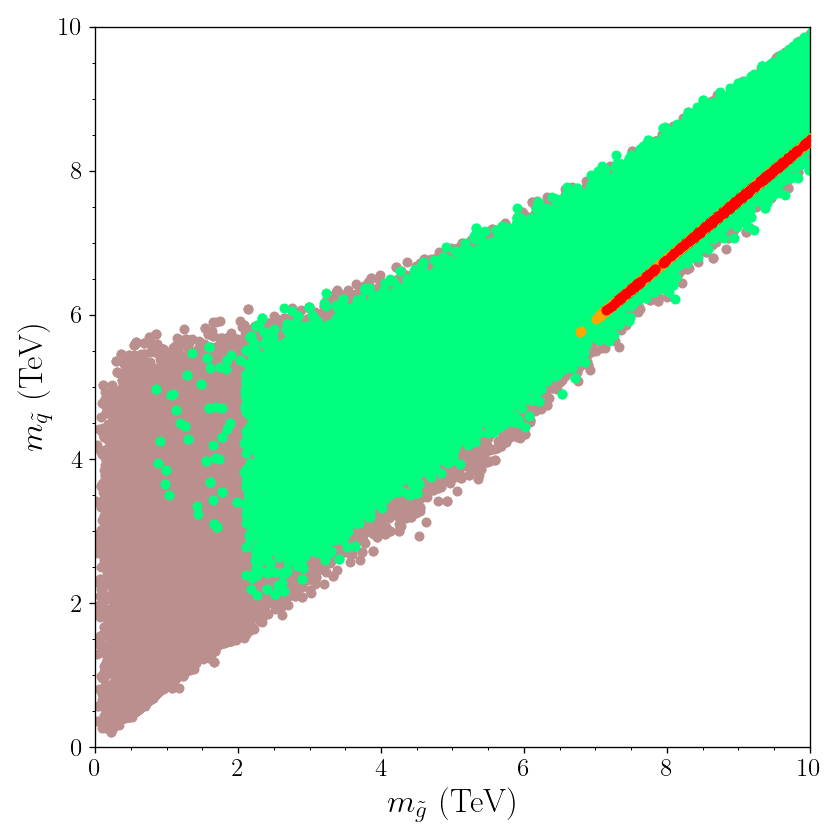}}%
\subfigure{\includegraphics[scale=0.4]{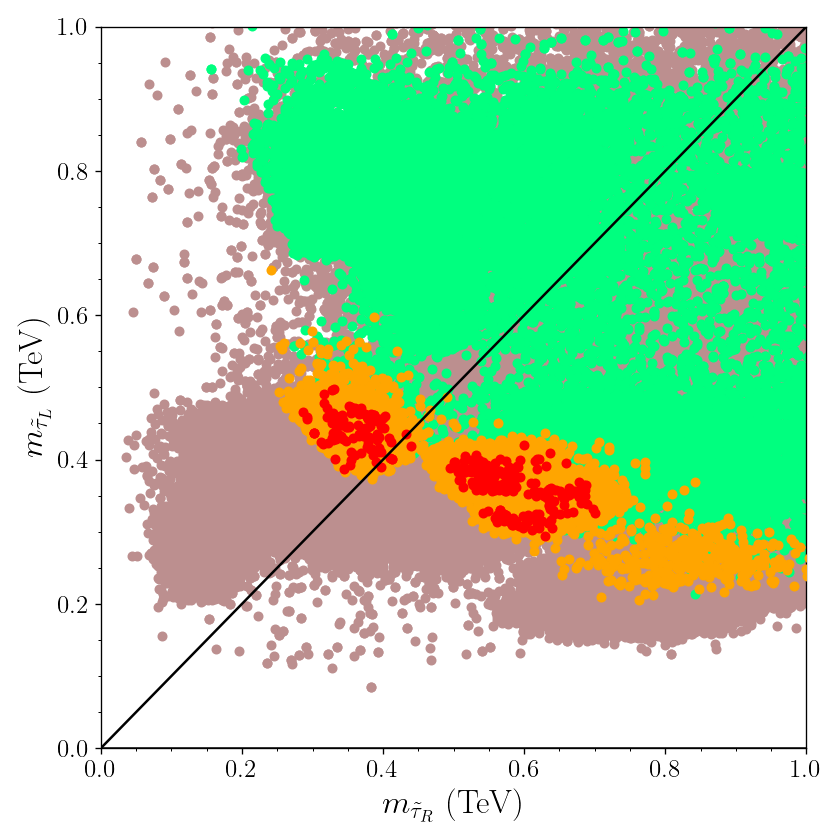}}
\subfigure{\includegraphics[scale=0.4]{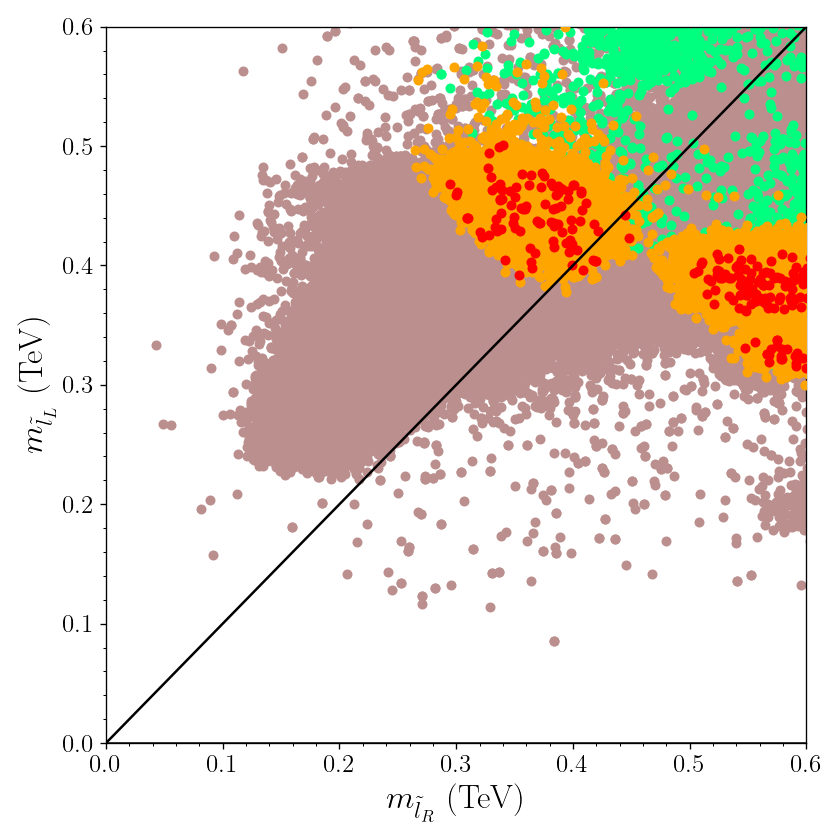}} %
\subfigure{\includegraphics[scale=0.4]{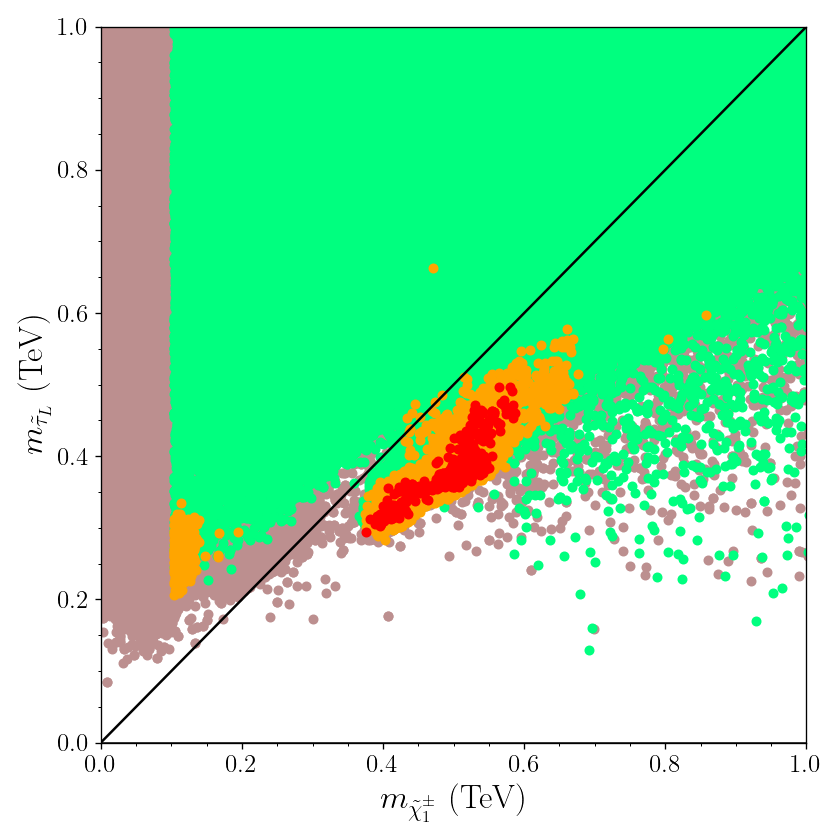}}

\caption{Sparticle spectrum with plots in the $m_{\tilde{q}}-m_{\tilde{g}}$, $m_{\tilde{\tau}_{L}}-m_{\tilde{\tau}_{R}}$, $m_{\tilde{\mu}_{L}}-m_{\tilde{\mu}_{R}}$ and $m_{\tilde{\tau}_{L}}-m_{\tilde{\chi}_{1}^{\pm}}$ planes. The color coding is the same as used in Figure \ref{fig:metastability}. The diagonal lines represent the degeneracy between the plotted masses.}
\label{fig:spectrum}
\end{figure}

We first discuss the impact of the muon $g-2$ measurements on the sparticle spectrum. As previously discussed, the supersymmetric muon $g-2$ contributions favor a relatively light spectrum which can be directly seen from the fundamental parameters shown in Figure \ref{fig:scalargauginos} with plots in the $\Delta a_{\mu}-m_{L}$, $\Delta a_{\mu}-m_{R}$, $\Delta a_{\mu}-M_{1}$ and $\Delta a_{\mu}-M_{2}$ planes. All solutions are compatible with the REWSB and LSP neutralino conditions. The green points are allowed by the mass bounds and constraints from rare $B-$meson decays. The red points form a subset of green and they satisfy the Planck bound on the relic density of LSP neutralino within $5\sigma$. The horizontal solid, dashed and dotted lines bound the regions which accommodate the muon $g-2$ solution within $1\sigma$, $2\sigma$ and $3\sigma$ respectively. The top panels in Figure \ref{fig:scalargauginos} reveal a strong impact on the SSB masses for the matter scalars, which mostly restricts $m_{L}$ to be lighter than about 600 GeV in order to resolve the muon $g-2$ discrepancy within $2\sigma$, whereas $m_{R}$ can be as heavy as about 1 TeV (green points). If one strictly requires these solutions to satisfy the correct relic density for the LSP neutralino (red points), these bounds are lowered further, namely $m_{L} \lesssim 400$ GeV and $m_{R} \lesssim 700$ GeV. A similar impact can also be observed on the SSB gaugino mass terms as shown in the bottom panels of Figure \ref{fig:scalargauginos}. The mass bounds and constraints from rare $B-$meson decays (green points) together with the muon $g-2$ solution within $2\sigma$ allow $M_{1}$ to be as heavy as about 2 TeV, while the Planck bound on the relic density of the LSP neutralino (red points) lower its bound to about 1 TeV. Our scans also result in a similar bound on $M_{2}$, which cannot be heavier than about 1 TeV.

Among these SSB mass parameters, $m_{R}$ and $M_{1}$ are determined through the LR breaking parameters $x_{LR}$ and $y_{LR}$ plotted in the $\Delta a_{\mu}-x_{LR}$ and $\Delta a_{\mu}-y_{LR}$ planes of Figure \ref{fig:damuLRbreaking}. The meanings of colors and horizontal lines are the same as in Figure \ref{fig:scalargauginos}. The LR breaking in the scalar sector can be as large as $x_{LR}\simeq 2.7$ consistent with the DM constraint, and the muon $g-2$ solutions favor the regions with $x_{LR}\simeq 0.7 - 1.1$. On the other hand, the impact of LR breaking in the gaugino sector strongly restricts $y_{LR} \lesssim -2$. Indeed, this arises mostly from the severe bound on the gluino mass, which can be satisfied with $M_{3} \gg M_{2}$. According to Eq.(\ref{eq:PSgauginos}), a large and negative $y_{LR}$ is required to realize $M_{1} \lesssim 1$ TeV and resolve the muon $g-2$ discrepancy to within $2\sigma$.

We display our results for the SSB trilinear scalar interaction term and $\tan\beta$ in the top panels of Figure \ref{fig:damuhiggs}, with the colors and horizontal lines defined as in Figure \ref{fig:scalargauginos}. The $\Delta a_{\mu}-A_{0}/m_{L}$ plane shows that the supersymmetric contributions to muon $g-2$  favor mostly negative $A_{0}$. Even though its ratio to $m_{L}$ can be as small as about 0.2, the relic density constraint from the Planck satellite requires $A_{0}/m_{L} \lesssim -1.2$, so that $|A_{0}| > m_{L}$. An interesting dependence on the supersymemtric muon $g-2$ contributions on $\tan\beta$ can be seen in the $\Delta a_{\mu}-\tan\beta$ plane, the mass bounds and consistent $B-$meson decays (green) can allow muon $g-2$ solutions within $2\sigma$ only in the regions with $\tan\beta \lesssim 22$. Requiring the LSP to be a consistent DM candidate (red points) lowers this bound further to $\tan\beta \lesssim 17$. This observation yields a strong tension between the experimental constraints and supersymmetric muon $g-2$ contributions, since the latter  are enhanced with large $\tan\beta$. Even though we present the results on the parameter $\tan\beta$ in terms of consistency with all the constraints employed in our analyses, the main negative impact comes from the constraint on the Higgs boson mass, which is shown in the bottom-left panel of Figure \ref{fig:damuhiggs}. 

As previously discussed, the loop contributions to the Higgs boson mass from stops are suppressed for larger $\tan\beta$ values, and a relatively small $A_{0}$ cannot compensate this suppression. 
Although it is possible to find models compatible with muon $g-2$ within $2\sigma$ and a Higgs boson mass close to 125 GeV, most of the solutions favored by muon $g-2$ predict the Higgs mass below the experimental value. Despite the enhancement arising from the sbottom and stau as discussed in Eq.(\ref{eq:higgscor}), for large $\tan\beta$ values, these contributions are relatively minor due to the restriction on $\mu\tan\beta$ from the vacuum  stability requirement on the scalar potential \cite{Carena:2012mw,Hisano:2010re,Kitahara:2013lfa}. Therefore, only models with a heavy stop can compensate the suppressed loop contribution to the Higgs boson mass. However, since the SUSY contribution to the muon $g-2$ problem restricts the SSB mass parameters, $m_{L}~(m_{R}) \lesssim 400~(1200)$ GeV, the low scale sparticle spectrum cannot involve such heavy stops. As a consequence, to solve the muon $g-2$ anomaly a combination of large $\tan\beta$ solutions with relatively large value of $\mu$ is needed. As shown in the $\Delta a_{\mu}-\mu$ plane in Figure \ref{fig:damuhiggs}, the desired solutions are realized for $\mu \gtrsim 3.5$ TeV, but the magnitude of $\mu$ is also limited since it cannot exceed 6 TeV. Realizing {such} large values of $\mu$ needs larger ranges for the SSB masses at the GUT scale for the gauginos and/or supersymmetric scalar particles, {which leads to a} heavier mass spectrum in the electroweak sector. In this case, the loop suppression encoded in the function $F_{{\rm N}}$ in Eq.(\ref{eq:binocont}) dominates over the enhancement from large $\mu$ values.

Although the constraints mentioned above significantly shrink the region compatible with muon $g-2$, the solutions that survive these restrictions remain viable after further considerations. For instance, the relatively small values for $\tan\beta$ and $A_{0}$, and mildly large values for $\mu$ will help to find stable vacua in this region for the scalar potential. Figure \ref{fig:metastability} displays our results relevant to the Higgs boson mass and metastability condition on the vacuum with plots in the $m_{L}-\tan\beta$, $m_{R}-\tan\beta$, $\mu-\tan\beta$ and $\mu\tan\beta^{{\rm max}}-\mu\tan\beta$ planes. All points are compatible with  REWSB and LSP neutralino conditions. Green points are allowed by the mass bounds and constraints from rare $B-$meson decays. The orange points form a subset of green, and they predict muon $g-2$ values within $2\sigma$. The red points as a subset of orange, additionally yield the correct relic density of the LSP neutralino measured by the Planck satellite within $5\sigma$ uncertainty. $\mu\tan\beta^{{\rm max}}$ represents the possible maximum values of $\mu\tan\beta$ allowed by the metastability condition on the scalar potential, and the diagonal line in the $\mu\tan\beta^{{\rm max}}-\mu\tan\beta$ plane indicates the solutions for which $\mu\tan\beta = \mu\tan\beta^{{\rm max}}$. The top panels show the ranges of $m_{L}$ and $m_{R}$ favored by the muon $g-2$ solution within $2\sigma$ as $m_{L} \lesssim 400$ GeV and $m_{R}\lesssim 700$ GeV as stated before, and these ranges can yield a consistent Higgs boson mass if $\tan\beta \lesssim 17$. The $\mu-\tan\beta$ plane shows that the allowed values of $\mu$ lie between about $3.5-5.5$ TeV, and its values tend to decrease with increasing $\tan\beta$. These  lead to $\mu\tan\beta$ values only as large as about 100 TeV as shown in the $\mu\tan\beta^{{\rm max}}-\mu\tan\beta$ plane, where $\mu\tan\beta^{{\rm max}}$ is obtained by applying the metastability condition on the vacuum \cite{Kitahara:2013lfa} to our data. As seen by comparing the solutions with respect to the diagonal line, only a few solutions that violate the metastability condition lie below the diagonal line (orange points). However, these solutions are already excluded by the Planck bound on the DM  relic density. The relic density constraint within $5\sigma$ bounds $\mu\tan\beta$ at about 70 TeV from above, and this is also more or less the bound imposed by the metastability condition. We should note here that we do not consider the quartic $|\tilde{\tau}_{L}\tilde{\tau}_{R}|^{2}$ term whose effects help to stabilize the vacuum and extend the metastability bound on $\mu\tan\beta$ to larger values \cite{Carena:2012mw}. Consequently, the solutions excluded by the metastability condition can still be available in further treatments on the vacuum stability.

In addition to the constraints discussed so far, the light sparticle spectrum has also been under active investigations by the current collider experiments at the LHC. We display some of these particles in Figure \ref{fig:spectrum}, and the impact  from LHC is expected to be improved during its Run3 phase, with plots in the $m_{\tilde{q}}-m_{\tilde{g}}$, $m_{\tilde{\tau}_{L}}-m_{\tilde{\tau}_{R}}$, $m_{\tilde{\mu}_{L}}-m_{\tilde{\mu}_{R}}$ and $m_{\tilde{\tau}_{L}}-m_{\tilde{\chi}_{1}^{\pm}}$ planes. The color coding is the same as used in Figure \ref{fig:metastability}, and the diagonal lines represent the degeneracy between the plotted masses. Even though we ensure the consistency with the bound on the gluino mass, the squarks might still happen to be inconsistently light due to the low values of $m_{L}$ and $m_{R}$. The $m_{\tilde{q}}-m_{\tilde{g}}$ plane, where $\tilde{q}$ denotes the lightest squark in the first two families, shows that the gluino cannot be lighter than about 6 TeV. Such heavy gluinos are somewhat beyond the reach of the current and near future collider experiments, but they might become testable in collisions with a higher center of mass energy and luminosity \cite{Altin:2019veq}. Such heavy gluinos can also enhance the squark masses by contributing through RGEs, and results show that the squarks are always heavier than about 5 TeV in the muon $g-2$ compatible region. 

In contrast to the colored particles as shown in the $m_{\tilde{\tau}_{L}}-m_{\tilde{\tau}_{R}}$ plane, the staus cannot be heavier than about 500 GeV. Similar mass scales can be observed for the first two family sleptons in the $m_{\tilde{\mu}_{L}}-m_{\tilde{\mu}_{R}}$ plane. We display our results in the flavor basis for these particles because the LHC analyses are able to provide sensitive results for sleptons in these mass scales, through the chargino and neutralino decays involving the sleptons \cite{CMS:2017fdz,Sirunyan:2017zss,Aaboud:2018jiw}. Such analyses can probe the first two family sleptons up to about 350 GeV, if they are mostly left-handed, while the chargino mass can be tested up to about 600 GeV in SUSY GUTs \cite{Shafi:2021jcg}. These bounds are expected to be lower when the stau is involved due to hadronic decays of $\tau-$leptons in the final state. In addition, the $m_{\tilde{\tau}_{L}}-m_{\tilde{\chi}_{1}^{\pm}}$ plane shows that the left-handed stau is always heavier than the chargino, which is expected to be Wino-like. In this case, the chargino decay into the light stau state is suppressed due to $SU(2)_{L}$ chirality. 

\section{Coannihilation Scenarios and DM Implications}
\label{sec:DM}

\begin{figure}[ht!]
\centering
\subfigure{\includegraphics[scale=0.4]{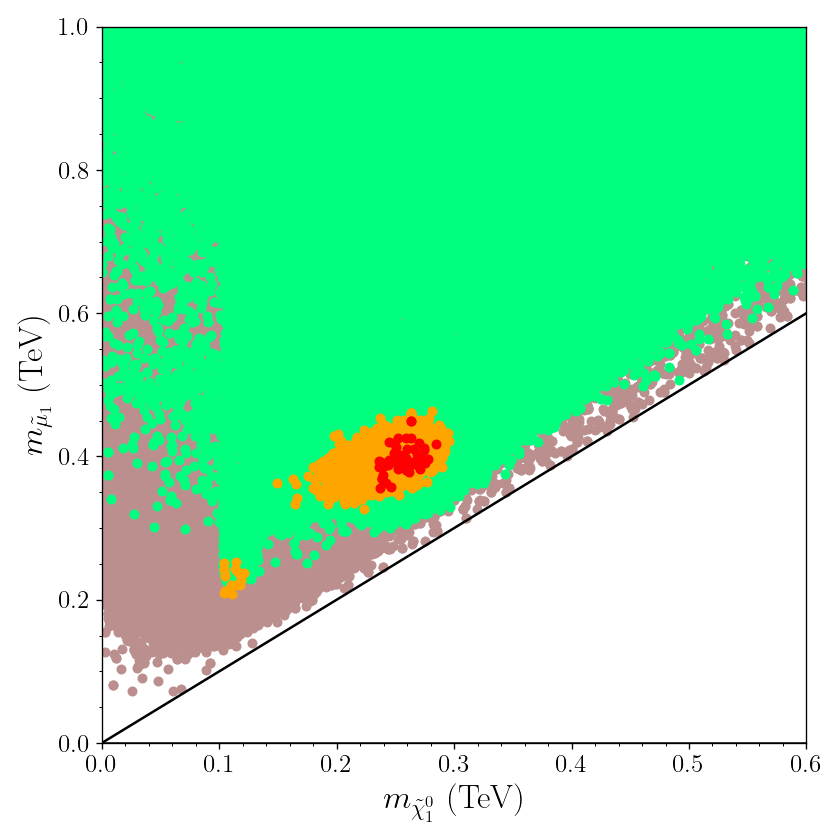}}%
\subfigure{\includegraphics[scale=0.4]{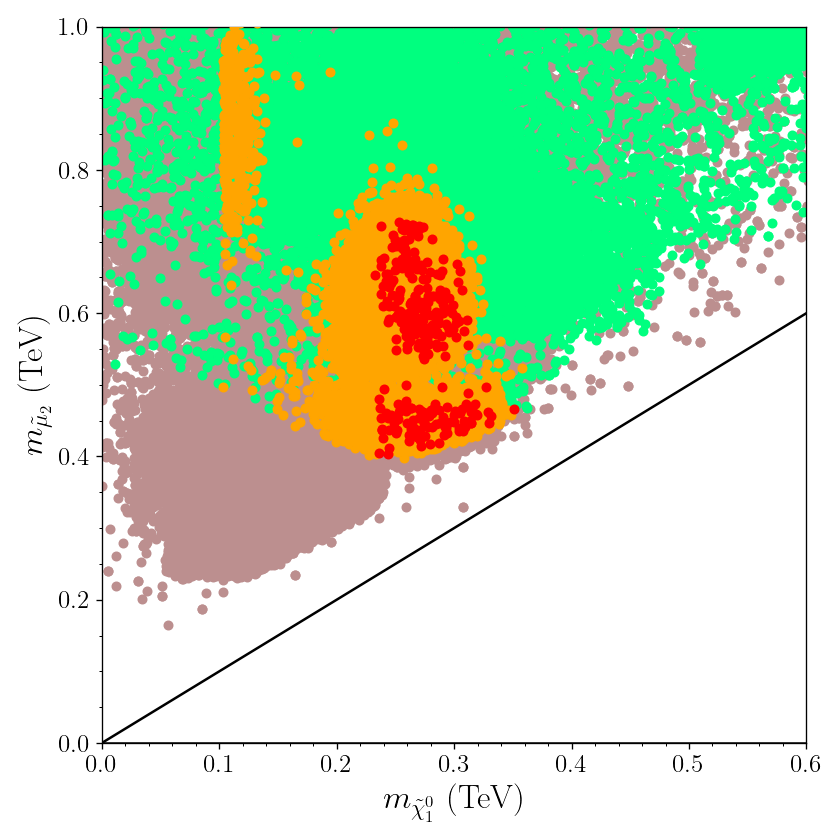}}
\subfigure{\includegraphics[scale=0.4]{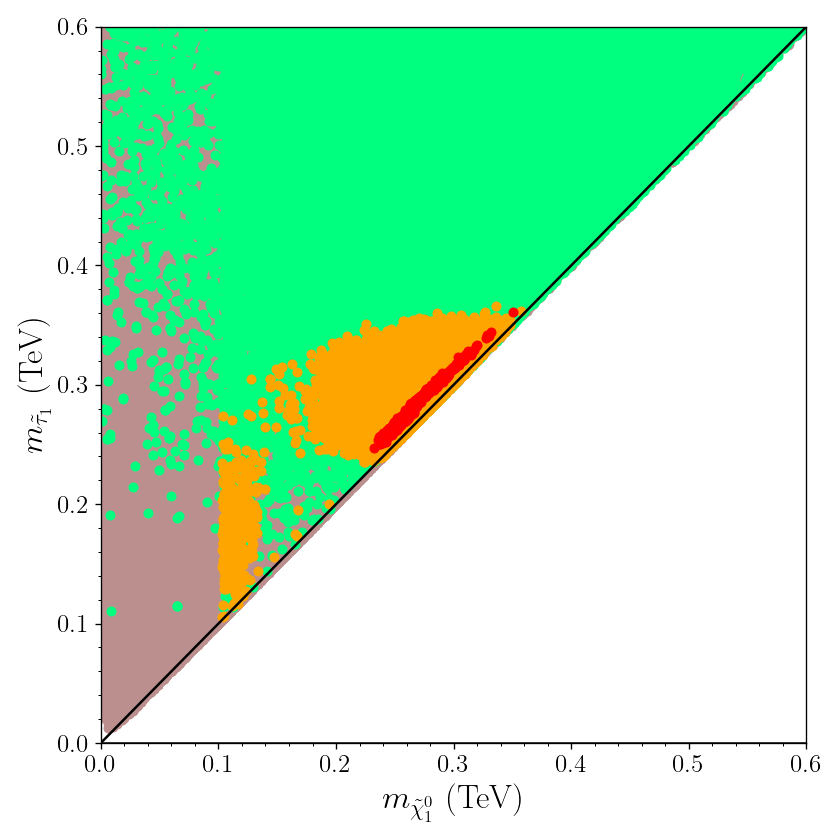}}%
\subfigure{\includegraphics[scale=0.4]{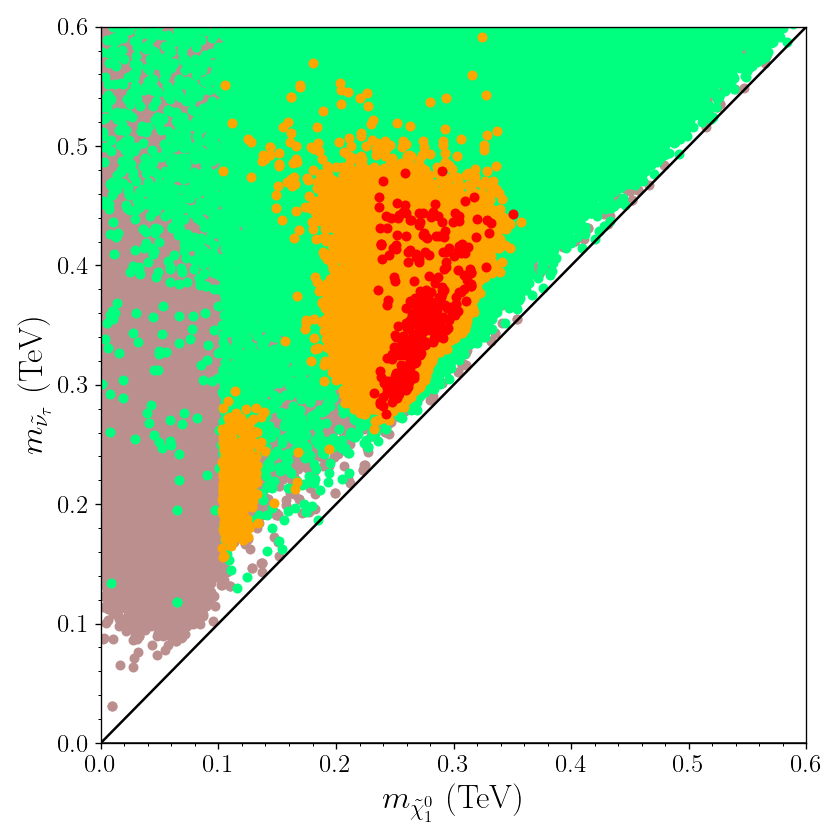}}
\caption{Slepton spectrum with respect to the LSP neutralino mass in the $m_{\tilde{\mu}_{1}}-m_{\tilde{\chi}_{1}^{0}}$, $m_{\tilde{\mu}_{2}}-m_{\tilde{\chi}_{1}^{0}}$, $m_{\tilde{\tau}_{1}}-m_{\tilde{\chi}_{1}^{0}}$, $m_{\tilde{\nu}_{1}}-m_{\tilde{\chi}_{1}^{0}}$ planes. The color coding is the same as in Figure \ref{fig:metastability}. The diagonal lines show the mass degeneracy between the plotted SUSY particles.}
\label{fig:sleptonsneut}
\end{figure}

\begin{figure}[ht!]
\centering
\subfigure{\includegraphics[scale=0.4]{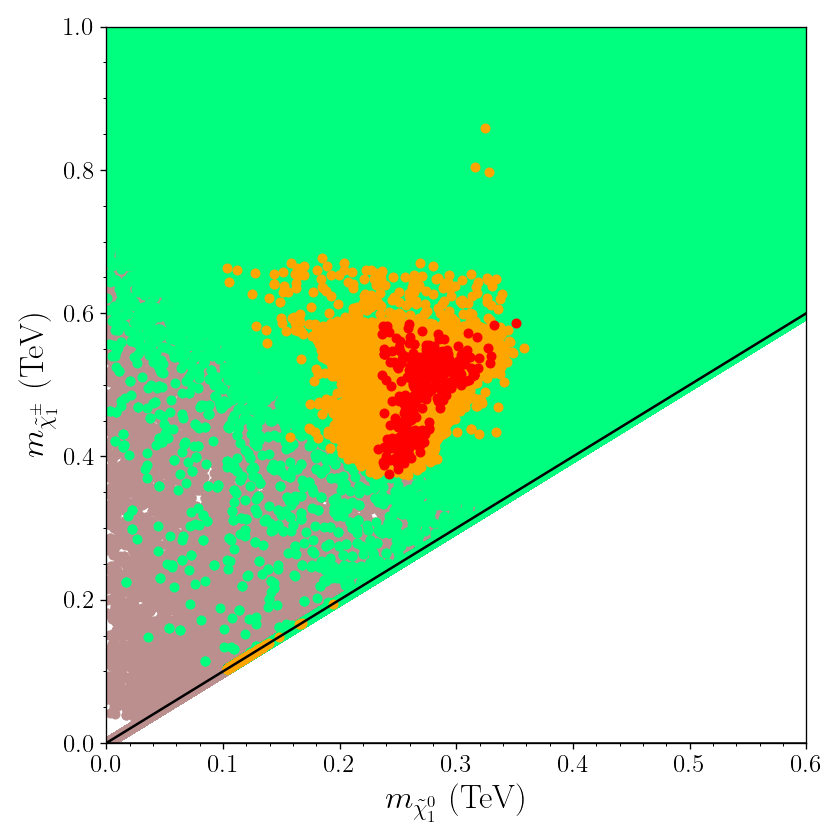}}%
\subfigure{\includegraphics[scale=0.4]{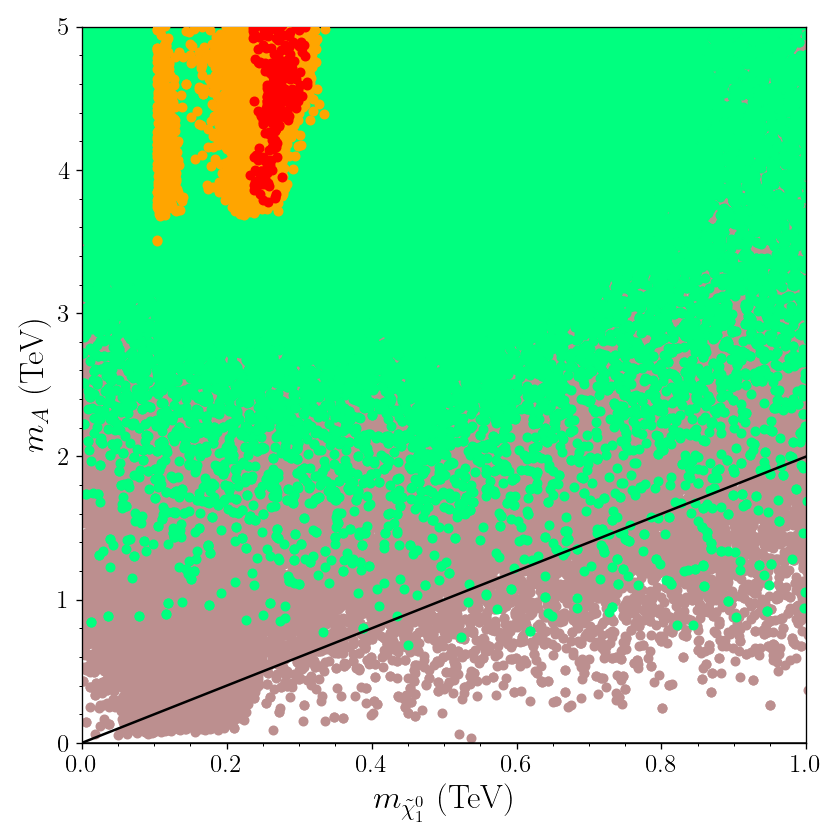}}
\caption{Slepton spectrum with respect to the LSP neutralino mass in the $m_{\tilde{\chi}_{1}^{\pm}}-m_{\tilde{\chi}_{1}^{0}}$ and $m_{A}-m_{\tilde{\chi}_{1}^{0}}$ planes. The color coding is the same as in Figure \ref{fig:metastability}. The diagonal line in the left plane represents the solutions with $m_{\tilde{\chi}_{1}^{\pm}}=m_{\tilde{\chi}_{1}^{0}}$, and that in the right plane indicates the $A-$resonance solutions with $m_{A}=2m_{\tilde{\chi}_{1}^{0}}$.}
\label{fig:coan2}
\end{figure}

In principle, the MSSM offers a variety of neutralino types - i.e. Bino, Wino and Higgsinos, and 
among them, the Higgsino can be found in the direct detection experiments because of its large scattering cross-sections with nuclei. However, the Planck measurements constrain its mass at about 700 GeV from below, and the current results of the direct detection experiments shift this bound even further to about 1 TeV (see, for instance, \cite{Raza:2018jnh}). Similar results also hold for Wino-like DM as well, in which case the SUSY contributions to muon $g-2$ will be suppressed. In this context, the Bino can still provide viable DM solutions since its scattering cross-section is lower by a few orders of magnitude, which may become accessible in the ongoing and future direct detection experiments. 

\begin{figure}[ht!]
\centering
\subfigure{\includegraphics[scale=0.4]{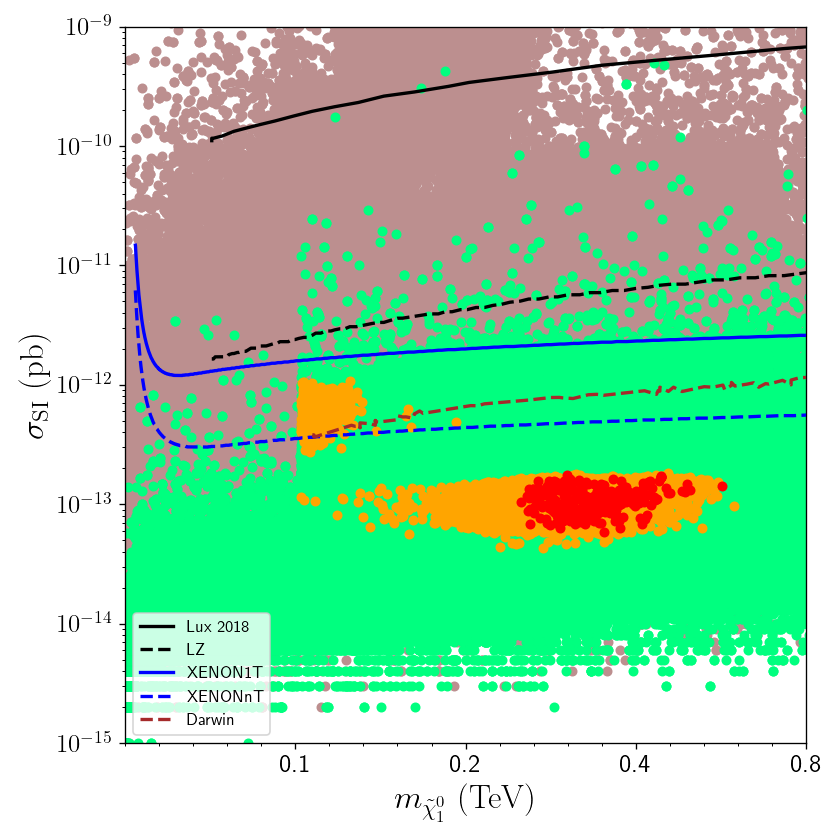}}%
\subfigure{\includegraphics[scale=0.4]{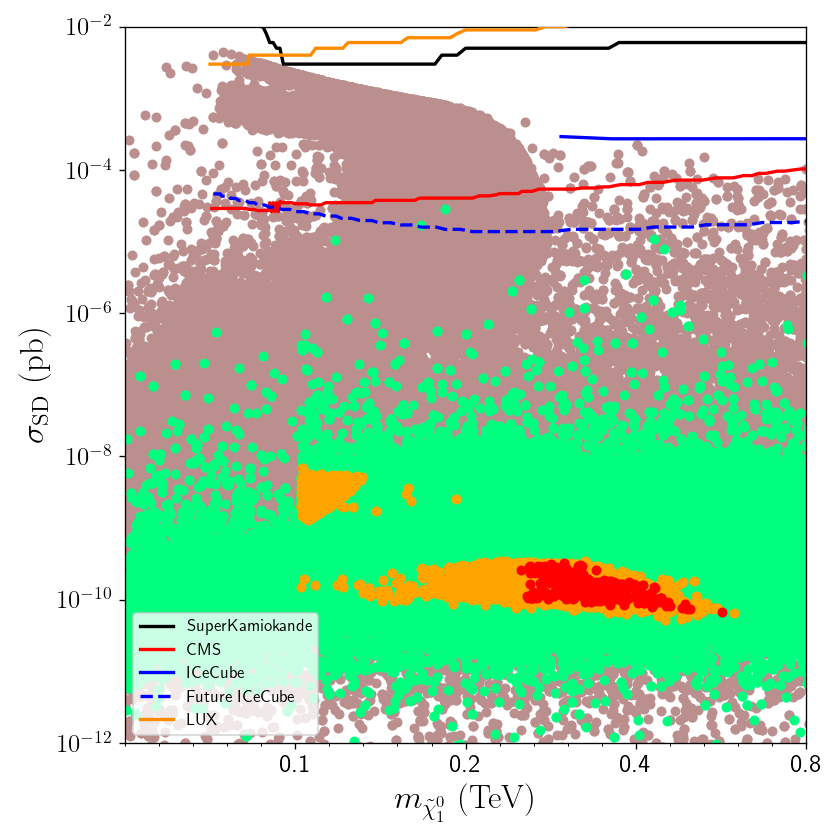}}
\caption{Spin-independent (left) and spin-dependent (right) scattering cross-section in correlation with the LSP neutralino mass. The color coding is the same as in Figure \ref{fig:metastability}. The curves represent the current and projected exclusion curves from several direct detection DM experiments, whose color coding is given in the legend for each plane. The current exclusions are represented by the solid curves, and the dashed curves display the projected { experimental sensitivity.}}
\label{fig:SISD}
\end{figure}

Since the muon $g-2$ compatible spectra involve relatively light sleptons, one can expect them to take part in the coannihilation scenarios. We display the masses of these sleptons and the LSP neutralino mass in Figure \ref{fig:sleptonsneut} with plots in the $m_{\tilde{\mu}_{1}}-m_{\tilde{\chi}_{1}^{0}}$, $m_{\tilde{\mu}_{2}}-m_{\tilde{\chi}_{1}^{0}}$, $m_{\tilde{\tau}_{1}}-m_{\tilde{\chi}_{1}^{0}}$, $m_{\tilde{\nu}_{1}}-m_{\tilde{\chi}_{1}^{0}}$ planes. The color coding is the same as in Figure \ref{fig:metastability}. The diagonal lines show the mass degeneracy between the plotted SUSY particles. The main restriction from muon $g-2$ is on the smuon masses, and the desired region is realized for $m_{\tilde{\mu}_{1}} \lesssim 440$ GeV and $m_{\tilde{\mu}_{2}} \lesssim 800$ with  the correct LSP relic density (red points), as seen in the top panels of Figure \ref{fig:sleptonsneut}. However, the mass differences between the LSP neutralino and smuons are larger than about 100 GeV, and thus they do not play a significant role in coannihilations. The mass differences between the LSP neutralino and staus are expected to be lower, and the plot in the $m_{\tilde{\tau}_{1}}-m_{\tilde{\chi}_{1}^{0}}$ plane shows that most of the solutions allowed by the Planck bound and muon $g-2$ accumulate around the diagonal line where $m_{\tilde{\tau}_{1}} \simeq m_{\tilde{\chi}_{1}^0}$. The results show that the stau-neutralino coannihilation scenario can be realized for staus as light as about 100 GeV, but the Planck bound on the relic density of LSP neutralino excludes the solutions with $m_{\tilde{\tau}_{1}} \lesssim 250$ GeV. The muon $g-2$ solution together with the Planck bound restricts the stau mass at about 350 GeV from above. This region also yields a sneutrino-neutralino coannihilation scenario for $m_{\tilde{\nu}_{\tau}} \gtrsim 250 $ GeV, as shown in the $m_{\tilde{\nu}_{1}}-m_{\tilde{\chi}_{1}^{0}}$ plane. However, these coannihilations {take} part up to about 350 GeV sneutrino mass, since otherwise the mass difference between the sneutrino and LSP is more than $10\%$ of the LSP neutralino mass.

In addition to the stau and sneutrino, we also display our results for the chargino-neutralino coannihilation scenario and $A-$resonance solutions in Figure \ref{fig:coan2} with plots in the 
$m_{\tilde{\chi}_{1}^{\pm}}-m_{\tilde{\chi}_{1}^{0}}$ and $m_{A}-m_{\tilde{\chi}_{1}^{0}}$ planes. The color coding is the same as in Figure \ref{fig:metastability}. The diagonal line in the left plane represents the solutions with $m_{\tilde{\chi}_{1}^{\pm}}=m_{\tilde{\chi}_{1}^{0}}$, and in the right plane it indicates the $A-$resonance solutions with $m_{A}=2m_{\tilde{\chi}_{1}^{0}}$. The results in the $m_{\tilde{\chi}_{1}^{\pm}}-m_{\tilde{\chi}_{1}^{0}}$ plane shows no solution for the chargino-neutralino coannihilation scenario, if the LSP neutralino itself is the exclusive DM candidate with the correct relic density (red points). The chargino mass can lie between about 400 and 600 GeV in this region. However, we also realize another region between about 100 and 200 GeV chargino mass, in which the chargino and LSP neutralino have almost degenerate masses. These solutions yield a considerable decrease in the relic density of LSP neutralino through chargino-neutralino coannihilation scenario. Comparing with the results displayed in the $m_{\tilde{\tau}_{1}}-m_{\tilde{\chi}_{1}^{0}}$ of Figure \ref{fig:sleptonsneut}, this region identifies chargino-neutralino and stau-neutralino coannihilation solutions simultaneously, which results in a very low relic density for the LSP neutralino that is incompatible with the Planck data. These solutions can be viable in scenarios which also include additional DM particles~\cite{Baer:2012by,Li:2014xqa}. Finally, we also display our results in the $m_{A}-m_{\tilde{\chi}_{1}^{0}}$ where the diagonal line shows the the regions of $A-$resonance solutions. The CP-odd Higgs boson cannot be lighter than about 3.5 TeV in the spectra compatible with muon $g-2$, which is quite heavy for the $A-$resonance solutions.

In addition to the Planck bound on the relic density of the LSP neutralino, the solutions discussed above can be also probed in the direct detection experiments. We display our results in Figure \ref{fig:SISD} for the spin-independent (left) and spin-dependent (right) scattering cross-sections versus the LSP neutralino mass. The color coding is the same as in Figure \ref{fig:metastability}. We also display the current and projected exclusion curves from several experiments such as LUX, LZ \cite{Akerib:2018lyp}, XENON \cite{Aprile:2020vtw} and DARWIN \cite{Aalbers:2016jon} for the spin-independent scattering cross-sections. The curves corresponding to the results of these experiments are listed in the legend of the $\sigma_{{\rm SI}}-m_{\tilde{\chi}_{1}^{0}}$ plane. The current exclusion {limits} are represented by the solid curves, and the dashed curves display the projected {experimental sensitivity}. As discussed before, the Bino-like DM typically yields low scattering cross-sections, and the $\sigma_{{\rm SI}}-m_{\tilde{\chi}_{1}^{0}}$ plane shows that the solutions lie slightly below the most sensitive projection, provided by XENONnT, and one may expect these solutions to be probed in near future. There is also an orange region between the current (dashed blue) and future projection (solid blue) curves of XENONnT. As stated before, these solutions can be available in non-standard DM scenarios and the analyses may soon reveal some results probing these solutions. We also present our results for the case of spin-dependent scattering of LSP neutralino in the $\sigma_{{\rm SD}}-m_{\tilde{\chi}_{1}^{0}}$ plane together with some experimental results from SuperKamiokande \cite{Tanaka:2011uf}, colliders \cite{Khachatryan:2014rra}, IceCube \cite{Abbasi:2009uz} and LUX \cite{Akerib:2016lao} experiments. Our results indicate that some further upgrades in such experiments should be able to probe the solutions compatible with the latest muon $g-2$ experimental results.

\begin{table}[ht!]
\centering
\setstretch{1.2}
\scalebox{0.8}{
\begin{tabular}{|c|cccccc|}
 \hline
  &Point 1&Point 2 &Point 3 &Point 4 &Point 5 &Point 6 \\ 
  \hline
  $m_{\widetilde{L}}$  & 335.319 & 375.605 & 311.513 & 377.107 & 330.994 & 361.925 \\ 
  
  $M_1$ & 616.148 & 672.019 & 633.965 & 767.804 & 787.614 & 1282.26 \\
  
  $M_{2L}$ & 731.027 & 747.131 & 603.037 & 721.392 & 698.554 & 327.194 \\
  
  $M_3$ & 4819.28 & 4920.39 & 4274.83 & 5067.11 & 4789.29 & 4505.82 \\
  
  $A_{0}/m_{\widetilde{L}}$ & -1.87 & -1.71 & -2.93 & -1.52 & -1.52 & 0.73 \\
  
  $\tan\beta$ & 9.08099 & 11.7627 & 16.1768 & 9.44659 & 7.63402 & 18.6131 \\
  
  \hline
  $x_{LR}$ & 0.79 & 0.88 & 2.12 & 0.88 & 0.86 & 2.2 \\
  
  $y_{LR}$ & -2.99 & -2.89 & -2.97 & -2.91 & -2.69 & -2.65 \\
  
  $m_{\widetilde{R}}$ & 265.065 & 329.393 & 660.442 & 331.076 & 285.988 & 795.804 \\
  
  $M_{2R}$ &-2185.94 & -2160.23 & -1793.28 & -2098.4 & -1880.17 & -866.78\\
  
  \hline
  $\mu$ & 4875.83  & 4951.28 & 4402.88 & 5098.57 & 4839.6 & 4493.68 \\
  
  $\Delta a_{\mu} \times 10^{10}$ & \bf{24.28} & 19.8 & 15.05 & 16.99 & 14.91 & 13.92\\
  
  \hline
  $m_h$ & 123.07 & 123.142 & 123.031 & \bf{123.541} & 123.03 & 123.215 \\
  
  $m_H$ & 4999.35 & 5061.57 & 4404.11 & 5222.41 & 5062.4 & 4498.81\\
  
  $m_A$ & 4999.31 & 5061.55 & 4404.11 & 5222.38 & 5062.35 & 4498.81 \\
  
  $m_{H^{\pm}}$ & 4996.57  & 5058.82 & 4401.99 & 5219.5 & 5059.58 & 4496.11 \\
  
  \hline
  $m_{\tilde{\chi}_{1}^{0}}$,$m_{\tilde{\chi}_{2}^{0}}$ & {\color{red}238}, 543 & {\color{red}263}, 557& {\color{red}250}, 441& {\color{red}305} , 532& {\color{red}315}, 514& {\color{red}194}, 530\\
  
  $m_{\tilde{\chi}_{3}^{0}}$,$m_{\tilde{\chi}_{4}^{0}}$ & 4991,4992& 5068, 5069& 4504, 4505& 5219, 5220 & 4956, 4957& 4607, 4607\\
  
  $m_{\tilde{\chi}_{1}^{\pm}}$,$m_{\tilde{\chi}_{2}^{\pm}}$ & 544, 4992 & 557, 5069& 442, 4505& 532, 5220& 515, 4957& {\color{red}195}, 4607\\
  \hline
  $m_{\tilde{g}}$ & 9549 & 9741 & 8555 & 10017 & 9501 & 9032\\
  
  $m_{\tilde{u}_1}$,$m_{\tilde{u}_2}$ & 8049, 8050 & 8210, 8211& 7231, 7259& 8437, 8439& 8012, 8012& 7630, 7677\\
  
  $m_{\tilde{t}_1}$,$m_{\tilde{t}_2}$ & 6979, 7525& 7123, 7670& 6267, 6724& 7319, 7886& 6945, 7492& 6712, 7126\\
  
  \hline
  $m_{\tilde{d}_1}$,$m_{\tilde{d}_2}$ &8051, 8055& 8212, 8216& 7231, 7262 & 8438, 8444& 8012, 8018& 7630, 7678\\
  
  $m_{\tilde{b}_1}$,$m_{\tilde{b}_2}$ & 7518, 8031& 7662, 8177& 6713, 7196& 7878, 8417& 7484, 8001& 7117, 7598\\
  
  \hline
  $m_{\tilde{\nu}_{e,\mu}}$,$m_{\tilde{\nu}_{\tau}}$ & 430, 418 & 459, 438& 369, 331& 451, 437& 403, 394& 293, 246\\
  
  $m_{\tilde{e}_{1}, \tilde{\mu}_1}$,$m_{\tilde{e}_{2}, \tilde{\mu}_2}$ & 368, 437 & 447, 465& 377, 703& 445, 458& 410, 446& 303, 937\\
  
  $m_{\tilde{\tau}_1}$,$m_{\tilde{\tau}_2}$ & {\color{red}255}, 484& {\color{red}284}, 535& {\color{red}268}, 690& {\color{red}320}, 523& {\color{red}325}, 489& {\color{red}200}, 919\\
  
  \hline
  $\sigma_{SI}(pb)$ & $1.15 \times 10^{-13}$ & $7.7 \times 10^{-14}$ & $6.7 \times 10^{-14}$ & $1.05 \times 10^{-13}$ & $1.71 \times 10^{-13}$ & $4.86 \times 10^{-13}$ \\
  
  $\sigma_{SD}(pb)$ & $1.05 \times 10^{-10}$ & $1.04 \times 10^{-10}$ & $1.69 \times 10^{-10}$ & $8.97 \times 10^{-11}$ & $1.06 \times 10^{-10}$ & $2.58 \times 10^{-9}$\\
  
  $\Omega h^2$ & 0.118 & 0.119 & 0.116 & 0.115 & 0.115 & 0.001\\
  \hline
  
\end{tabular}}
\caption{Benchmark points satisfy the mass bounds, the constraints from rare B-meson decays, Planck bound within $5\sigma$ and muon $g-2$ solution within $2\sigma$ at most. All masses are in GeV. Point 1 shows a solution that has muon $g-2$ value closest to its world average. Points 2 and 3 display solutions with the possible largest $\tan\beta$ consistent with the muon $g-2$ solution within $1\sigma$ and $2\sigma$ respectively. Point 4 depicts a solution for the possible heaviest mass of the Higgs boson in the muon $g-2$ compatible region. Point 5 refers to solutions with the smallest difference in $m_{\tilde{\chi}_{1}^{0}}$ and $m_{\tilde{\tau}_1}$. Point 6 exemplifies a solution with nearly degenerate masses for the LSP neutralino, chargino and stau.}
\label{tab1}
\end{table}

Finally, we exemplify our findings with six benchmark points listed in Table \ref{tab1}. The points selected are consistent with the mass bounds, the constraints from rare B-meson decays, Planck bound within $5\sigma$ and muon $g-2$ solution within ${1\sigma-2\sigma}$. All masses are in GeV. Point 1 shows a solution that has muon $g-2$ value closest to its world average. Points 2 and 3 display solutions with the largest possible $\tan\beta$ consistent with the resolution of the muon $g-2$ anomaly within $1\sigma$ and $2\sigma$ respectively. Point 4 depicts a solution for the possible heaviest mass of the Higgs boson, and Point 5 refers to solutions with the smallest mass difference between $m_{\tilde{\chi}_{1}^{0}}$ and $m_{\tilde{\tau}_1}$. Point 6 exemplifies solutions for the chargino and LSP neutralino that are nearly degenerate in mass. The consistent relic density of LSP neutralino is achieved through the stau-neutralino and sneutrino-neutralino coannihilation scenarios as discussed before. All the points listed in Table \ref{tab1} except Point 6, exemplify spectra in which the stau happens to be the NLSP (shown in red), while the other slepton masses are lighter than about 450 GeV. On the other hand, the squarks mostly weigh more than about 7.2 TeV. The third family squarks can be slightly lighter ($m_{\tilde{t}_{1}}\gtrsim 5.6$ TeV), and the gluinos are heavier than about 8.5 TeV. Similarly, the muon $g-2$ solutions yield heavy Higgs bosons ($\gtrsim 4$ TeV) except the SM-like Higgs boson whose mass is around the edge allowed by the uncertainty. Even though they have spin-independent LSP neutralino scattering cross-sections lower than the current exclusion limits of the direct detection DM experiments, Points 1, 4, 5 and 6 predict a scattering cross-section on the order of $10^{-13}$ pb, which hopefully may be tested in the near future. {The first five benchmark points in Table \ref{tab:fund} display solutions with a Bino-like ($\gtrsim 99\%$) LSP neutralino, and Point 6 shows a Wino-like LSP neutralino, with a relic density well below the Planck bound. Despite the low relic density, the latter solution is viable if an additional DM candidate is also present \cite{Baer:2012by,Li:2014xqa}.}

\section{Conclusions}
\label{sec:conc}

We consider a class of SUSY GUTs in which the $SO(10)$ breaks via $4-2-2$ to MSSM. We assume the $SO(10)$ breaking to $4-2-2$ happens through the VEV of a Higgs fields in the $210$ dimensional representation that also breaks the LR symmetry. The GUT spectrum also involves $126$ and $\overline{126}$ dimensional Higgs representations, whose VEVs break $4-2-2$ to the MSSM gauge group. With such a breaking mechanism the possible $D-$term contributions to the SSB scalar masses are canceled, but the absence of LR symmetry yields different SSB masses at $\mgut$ for the left and right-handed supersymmetric scalars, as well as leading to non-universal gaugino masses.

We confront the predictions of this class of SUSY models with the recent Fermilab measurement of the anomalous muon $g-2$. We found a region of parameter space such that the model prediction agress within $1-2\sigma$ with experimental result. This region restricts the masses of sleptons, neutralino and chargino to be lighter than about 1 TeV and a typical spectrum involves NLSP staus whose mass has an upper limit of about 400 GeV. The Planck bound on the DM relic density restricts the stau mass from below as $m_{\tilde{\tau}}\gtrsim 250$ GeV. The masses of the first two charged slepton families are found to be slightly heavier, namely $m_{\tilde{l}}\gtrsim 350$ GeV, where $\tilde{l}=\tilde{e},\tilde{\mu}$. Even though the supersymmetric muon $g-2$ contributions depend on the chargino mass, values for the latter larger than about 900 GeV are still compatible with the experimental muon $g-2$ measurement within $2\sigma$. On the other hand, the DM relic density constraint imposes an upper bound of about 600 GeV for the chargino mass. Chargino and sleptons masses may be further restricted by the current and Run3 experiments at LHC. However, the dominant right-handed chirality of the lighter sleptons suppress chargino decays, which is mostly Wino-like, into these relatively light sleptons. 

In contrast to the sleptons, chargino and LSP neutralino, the other SUSY particles are heavy. The supersymmetric muon $g-2$ contributions are enhanced by the relatively large $\mu-$ term, and as a result, the Higgsino masses are of about 4 TeV or heavier, with similar mass scales for the additional scalars present in the MSSM. However, the mass prediction for the SM-like Higgs becomes low at large $\tan\beta$. This is due to the fact that the stop contributions to the Higgs boson mass decrease with $\tan\beta$, and a heavier SUSY mass spectrum is needed to compensate this suppression. Therefore, considering the $\tan\beta$ enhancement in the supersymmetric contributions to muon $g-2$, there exists a significant tension between the muon $g-2$ resolution and the Higgs boson mass constraint. We find that the two requirements cannot satisfy the experimental results if $\tan\beta \gtrsim 17$. We also discussed vacuum stability which is easily satisfied for low $\tan\beta$ values. Indeed, we find that a small portion of the parameter space that violates the metastability condition on the scalar potential is already excluded by the DM relic density constraint. 

The solution of the muon $g-2$ discrepancy within $2\sigma$ imposes an upper limit of about 350 GeV on the LSP neutralino mass. The LSP neutralino is pre-dominantly a Bino, with the lighter chargino mostly a Wino. Even though Bino-like DM typically yields a large relic density, its density can be lowered through coannihilation and annihilation processes, and the light sleptons in the spectrum are suitable for such coannihilation processes. The stau-neutralino coannihilation dominantes in reducing the LSP neutralino density, and snuetrino-neutralino coannihilation provides minor contributions for realizing the correct relic density of LSP neutralino. In this context, imposing the Planck bound on the relic abundance of LSP neutralino on the muon $g-2$ compatible region results in an upper bound on the stau mass, $m_{\tilde{\tau}_{1}} \lesssim 360$ GeV. Even though the current LHC analyses \cite{ATLAS:2018ojr,CMS:2018eqb} constrain severely the light slepton masses, the results from these analyses are crucially sensitive to the mass difference between the NLSP and LSP. If the mass difference is below 80 GeV \cite{CMS:2019zmn,Chakraborti:2020vjp}, the possible signal processes involve soft tau leptons in their final states, which significantly raise the uncertainty in the experimental analyses due to relatively lower sensitivity of the detectors to tau lepton and large QCD background \cite{CMS:2019eln}. In our model the mass difference between the stau and LSP neutralino is not greater than about 15 GeV in the stau-neutralino coannihilation region, and therefore the solutions we found remain intact. In addition, we identify a region with both chargino-neutralino and stau-neutralino coannihilations where the resultant relic density becomes lower than the current measurements of the Planck satellite.

Despite its low scattering cross-section, a Bino-like DM can provide compelling predictions since the current sensitivity of the direct DM detection experiments has improved significantly. Our solutions predict a spin-independent scattering cross-sections on the order of $10^{-13}$ pb, which is only slightly lower than the projected sensitivity of the XENON experiment and hopefully can be tested in the near future. We display results for the spin-dependent scattering of DM which lie well below the current and projected experimental sensitivities. We exemplify our findings using six benchmark points.   

\section*{Acknowledgment}
This work is supported in part by the United States Department of Energy grant DE-SC0013880 (QS and AT). The research of MEG and CSU is supported in part by the Spanish MICINN, under grant PID2019-107844GB-C22. We acknowledge Information Technologies (IT) resources at the University Of Delaware, specifically the high performance computing resources for the calculation of results presented in this paper.

\bibliographystyle{JHEP}
\bibliography{mybibtex}

\end{document}